\input harvmac.tex



\def\prop#1{\bigskip\noindent{\bf Proposition #1} }

\def\rmk#1{\bigskip\noindent{\bf Remarks} }


\def\unlockat{\catcode`\@=11}
\def\lockat{\catcode`\@=12}

\unlockat

\def\newsec#1{\global\advance\secno by1\message{(\the\secno. #1)}
\global\subsecno=0\global\subsubsecno=0\eqnres@t\noindent
{\bf\the\secno. #1}
\writetoca{{\secsym} {#1}}\par\nobreak\medskip\nobreak}
\global\newcount\subsecno \global\subsecno=0
\def\subsec#1{\global\advance\subsecno
by1\message{(\secsym\the\subsecno. #1)}
\ifnum\lastpenalty>9000\else\bigbreak\fi\global\subsubsecno=0
\noindent{\it\secsym\the\subsecno. #1}
\writetoca{\string\quad {\secsym\the\subsecno.} {#1}}
\par\nobreak\medskip\nobreak}
\global\newcount\subsubsecno \global\subsubsecno=0
\def\subsubsec#1{\global\advance\subsubsecno by1
\message{(\secsym\the\subsecno.\the\subsubsecno. #1)}
\ifnum\lastpenalty>9000\else\bigbreak\fi
\noindent\quad{\secsym\the\subsecno.\the\subsubsecno.}{#1}
\writetoca{\string\qquad{\secsym\the\subsecno.\the\subsubsecno.}{#1}}
\par\nobreak\medskip\nobreak}

\def\subsubseclab#1{\DefWarn#1\xdef
#1{\noexpand\hyperref{}{subsubsection}%
{\secsym\the\subsecno.\the\subsubsecno}%
{\secsym\the\subsecno.\the\subsubsecno}}%
\writedef{#1\leftbracket#1}\wrlabeL{#1=#1}}
\lockat

\input epsf

\def\figin{\epsfcheck\figin}\def\figins{\epsfcheck\figins}
\def\epsfcheck{\ifx\epsfbox\UnDeFiNeD
\message{(NO epsf.tex, FIGURES WILL BE IGNORED)}
\gdef\figin##1{\vskip2in}\gdef\figins##1{\hskip.5in}
\else\message{(FIGURES WILL BE INCLUDED)}%
\gdef\figin##1{##1}\gdef\figins##1{##1}\fi}
\def\DefWarn#1{}
\def\figinsert{\goodbreak\midinsert}
\def\ifig#1#2#3{\DefWarn#1\xdef#1{fig.~\the\figno}
\writedef{#1\leftbracket fig.\noexpand~\the\figno}%
\figinsert\figin{\centerline{#3}}\medskip\centerline{\vbox{\baselineskip12pt
\advance\hsize by -1truein\noindent\footnotefont{\bf Fig.~\the\figno:} #2}}
\bigskip\endinsert\global\advance\figno by1}

\noblackbox

\def\IL{\relax{\rm I\kern-.18em L}}
\def\IH{\relax{\rm I\kern-.18em H}}
\def\IR{\relax{\rm I\kern-.18em R}}
\def\IC{\relax\hbox{$\inbar\kern-.3em{\rm C}$}}
\def\IT{\relax\hbox{$\inbar\kern-.3em{\rm T}$}}
\def\IZ{\relax\ifmmode\mathchoice
{\hbox{\cmss Z\kern-.4em Z}}{\hbox{\cmss Z\kern-.4em Z}}
{\lower.9pt\hbox{\cmsss Z\kern-.4em Z}} {\lower1.2pt\hbox{\cmsss
Z\kern-.4em Z}}\else{\cmss Z\kern-.4em Z}\fi}

\def\CN {{\cal N}}
\def\CR {{\cal R}}
\def\CD {{\cal D}}

\def\CO {{\cal O}}

\def\CB {{\cal B}}
\def\CS {{\cal S}}
\def\CA{{\cal A}}


\def\CN {{\cal N}}

\def\CO {{\cal O}}

\def\CS {{\cal S }}
\def\CW {{\cal W }}

\font\manual=manfnt \def\dbend{\lower3.5pt\hbox{\manual\char127}}

\def\IZ{\relax\ifmmode\mathchoice
{\hbox{\cmss Z\kern-.4em Z}}{\hbox{\cmss Z\kern-.4em Z}}
{\lower.9pt\hbox{\cmsss Z\kern-.4em Z}} {\lower1.2pt\hbox{\cmsss
Z\kern-.4em Z}}\else{\cmss Z\kern-.4em Z}\fi}
\def\half {{1\over 2}}

\def\CN {{\cal N}}

\def\CO {{\cal O}}

\def\CS {{\cal S }}


\def\IZ{\relax\ifmmode\mathchoice
{\hbox{\cmss Z\kern-.4em Z}}{\hbox{\cmss Z\kern-.4em Z}}
{\lower.9pt\hbox{\cmsss Z\kern-.4em Z}} {\lower1.2pt\hbox{\cmsss
Z\kern-.4em Z}}\else{\cmss Z\kern-.4em Z}\fi}
\def\IB{\relax{\rm I\kern-.18em B}}
\def\IC{{\relax\hbox{$\inbar\kern-.3em{\rm C}$}}}
\def\ID{\relax{\rm I\kern-.18em D}}
\def\IE{\relax{\rm I\kern-.18em E}}
\def\IF{\relax{\rm I\kern-.18em F}}
\def\IG{\relax\hbox{$\inbar\kern-.3em{\rm G}$}}
\def\IGa{\relax\hbox{${\rm I}\kern-.18em\Gamma$}}
\def\IH{\relax{\rm I\kern-.18em H}}
\def\II{\relax{\rm I\kern-.18em I}}
\def\IK{\relax{\rm I\kern-.18em K}}
\def\IP{\relax{\rm I\kern-.18em P}}
\def\IQ{\relax\hbox{$\inbar\kern-.3em{\rm Q}$}}

\def\inbar{\,\vrule height1.5ex width.4pt depth0pt}

\def\rmk#1{\bigskip\noindent{\bf Remarks #1}}

\font\cmss=cmss10 \font\cmsss=cmss10 at 7pt
\def\IR{\relax{\rm I\kern-.18em R}}


\def\boxit#1{\vbox{\hrule\hbox{\vrule\kern8pt
\vbox{\hbox{\kern8pt}\hbox{\vbox{#1}}\hbox{\kern8pt}}
\kern8pt\vrule}\hrule}}
\def\mathboxit#1{\vbox{\hrule\hbox{\vrule\kern8pt\vbox{\kern8pt
\hbox{$\displaystyle #1$}\kern8pt}\kern8pt\vrule}\hrule}}


\def\inbar{\,\vrule height1.5ex width.4pt depth0pt}

\font\cmss=cmss10 \font\cmsss=cmss10 at 7pt
\def\IR{\relax{\rm I\kern-.18em R}}



\lref\bateman{A. Erdelyi et. al. , {\it Higher
Transcendental Functions, vol. I, Bateman manuscript project}
(1953) McGraw-Hill}

\lref\borchaut{R. Borcherds,
``Automorphic forms with singularities on Grassmannians,''
alg-geom/9609022}

\lref\donint{S.K. Donaldson, ``Connections,
Cohomology and the intersection forms of 4-manifolds,'' J. Diff.
Geom. {\bf 24 } (1986)275.}

\lref\finstern{R. Fintushel and R.J. Stern,
``The blowup formula for Donaldson invariants,'' alg-geom/9405002;
Ann. Math. {\bf 143} (1996) 529.}

\lref\mw{G. Moore and E. Witten, ``Integration over
the $u$-plane in Donaldson theory," hep-th/9709193; Adv. Theor.
Math. Phys. {\bf 1} (1998) 298.  }

\lref\swi{N. Seiberg and E. Witten,
``Electric-magnetic duality, monopole condensation, and confinement
in ${\cal N}=2$ supersymmetric Yang-Mills Theory,'' hep-th/9407087;
Nucl. Phys. {\bf B426} (1994) 19}

\lref\swii{N. Seiberg and E. Witten,
``Monopoles, duality and chiral symmetry breaking in ${\cal N}=2$
supersymmetric QCD,'' hep-th/9408099; Nucl. Phys. {\bf B431} (1994)
484. }

\lref\ad{P.C. Argyres and M.R. Douglas, ``New phenomena in $SU(3)$
supersymmetric gauge theory," hep-th/9505062; Nucl. Phys. {\bf
B448} (1995) 93.}
\lref\ds{M.R. Douglas and S.H. Shenker, ``Dynamics of $SU(N)$ supersymmetric
gauge theory," hep-th/9503163; Nucl. Phys. {\bf B447} (1995) 271.}

\lref\giveonkut{A. Giveon and D. Kutasov,
``Brane Dynamics and Gauge Theory,'' hep-th/9802067}

\lref\vw{C. Vafa and E. Witten,
``A strong coupling test of $S$-duality,'' hep-th/9408074; Nucl.
Phys. {\bf B431} (1994) 3.}

\lref\monopole{E. Witten, ``Monopoles and
four-manifolds,''  hep-th/9411102; Math. Res. Letters {\bf 1}
(1994) 769.}

\lref\witteni{E. Witten, ``On $S$-duality in abelian
gauge theory,'' hep-th/9505186; Selecta Mathematica {\bf 1} (1995)
383.}

\lref\wittk{E. Witten, ``Supersymmetric Yang-Mills theory
on a four-manifold,''  hep-th/9403193; J. Math. Phys. {\bf 35}
(1994) 5101.}

\lref\lns{A. Losev, N. Nekrasov, and S. Shatashvili, ``Issues in
topological gauge theory," hep-th/9711108; Nucl. Phys. {\bf B 534} (1998)
549. ``Testing Seiberg-Witten
solution," hep-th/9801061.}

\lref\ms{T. Masuda and H. Suzuki, ``On explicit evaluations around the
conformal point in ${\cal N}=2$ supersymmetric Yang-Mills
theories," hep-th/9612240; Nucl. Phys. {\bf B495} (1997) 149.}

\lref\mm{M. Mari\~no and G. Moore, ``Integrating over the Coulomb branch in
${\cal N}=2$ gauge theory," hep-th/9712062;  Nucl. Phys. Proc.
Suppl. {\bf b68} (1998) 336.}

\lref\matone{M. Matone, ``Instantons and recursion relations in ${\cal N}=2$
supersymmetric gauge theory," hep-th/9506102; Phys. Lett. {\bf
B357} (1995) 342.}

\lref\humphreys{J.E. Humphreys, {\it Introduction to Lie algebras and
representation theory}, Springer-Verlag, 1972.}

\lref\fm{R. Friedman and J.W. Morgan,
``Algebraic surfaces and Seiberg-Witten invariants,''
alg-geom/9502026; J. Alg. Geom. {\bf 6} (1997) 445. ``Obstruction
bundles, semiregularity, and Seiberg-Witten invariants'',
alg-geom/9509007.}

\lref\morganbk{J.W. Morgan, {\it The Seiberg-Witten equations and applications
to the topology of smooth four-manifolds}, Princeton University
Press, 1996.}

\lref\DoKro{S.K.~ Donaldson and P.B.~ Kronheimer,
{\it The Geometry of Four-Manifolds}, Clarendon Press, Oxford,
1990.}

\lref\fmbook{R. Friedman and J.W. Morgan,
{\it Smooth four-manifolds and complex surfaces}, Springer Verlag,
1991.}

\lref\tqft{E. Witten,
``Topological Quantum Field Theory,'' Commun. Math. Phys. {\bf 117}
(1988) 353.}

\lref\apsw{P.C. Argyres, M.R. Plesser, N. Seiberg, and E. Witten, ``New ${\cal
N}=2$ superconformal field theories in four dimensions,"
hep-th/9511154; Nucl. Phys. {\bf B461} (1996) 71.}

\lref\eguchisc{T. Eguchi, K. Hori, K. Ito, and S.K. Yang, ``Study of ${\cal
N}=2$ superconformal field theories in four dimensions,"
hep-th/9603002; Nucl. Phys. {\bf B471} (1996) 430.}

\lref\BSV{M. Bershadsky, V. Sadov, and
C. Vafa, ``D-Branes and Topological Field Theories,'' Nucl. Phys.
{\bf B463} (1996) 420; hep-th/9511222.}

\lref\nsc{M. Mari\~no and G. Moore, ``Donaldson invariants for nonsimply
connected manifolds," hep-th/9804104.}

\lref\munnsc{V. Mu\~noz, ``Basic classes for four-manifolds
not of simple type,'' math.DG/9811089.}

\lref\freedman{D. Anselmi, J. Erlich, D.Z. Freedman,
and A.A. Johansen, ``Nonperturbative formulas for central functions
of supersymmetric gauge theories,'' hep-th/9708042;
Nucl.Phys.B526:543-571,1998 ; ``Positivity constraints on anomalies
in supersymmetric gauge theories,'' hep-th/9711035;
Phys.Rev.D57:7570-7588,1998}

\lref \pidtyurin{V. Pidstrigach and A. Tyurin, ``Localization of the Donaldson
invariants along Seiberg-Witten classes,'' dg-ga/9507004.}

\lref\feehan{P.M.N. Feehan and  T.G. Leness, ``$PU(2)$ monopoles and relations
between four-manifold invariants," dg-ga/9709022; ``$PU(2)$
monopoles I: Regularity, Uhlenbeck compactness, and
transversality," dg-ga/9710032; ``$PU(2)$ monopoles II:
Highest-level singularities and realtions between four-manifold
invariants," dg-ga/9712005. }

\lref\baryon{J.M.F. Labastida and M. Mari\~no, ``Twisted baryon
number in ${\cal N}=2$ supersymmetric QCD," hep-th/9702054; Phys. Lett.
{\bf B 400} (1997) 323.}

\lref\seiberg{N. Seiberg, Phys. Lett. {\bf B 318} (1993) 469; Phys. Rev.
{\bf D 49} (1994) 6857.}
\lref\ils{K. Intriligator, R.G. Leigh and N. Seiberg, Phys. Rev. {\bf D 50}
(1994) 1052.}
\lref\is{K. Intriligator and N. Seiberg, ``Lectures on ${\cal N}=1$
supersymmetric gauge theories
and electric-magnetic duality," hep-th/9509066; Nucl. Phys. Proc.
Suppl. {\bf B45} (1996)1. }

\lref\gompfbook{R. E. Gompf and A.I. Stipsicz, {\it An introduction to
four-manifolds and
Kirby calculus}, available in R.E. Gompf's homepage at
http://www.ma.utexas.edu.}

\lref\bilalfer{A. Bilal and F. Ferrari, ``The BPS spectra and
superconformal points in massive
${\cal N}=2$ supersymmetric QCD", hep-th/9706145; Nucl. Phys. {\bf
B516} (1998) 175.}

\lref\ky{H. Kanno and S.-K. Yang, ``Donaldson-Witten function of
massless ${\cal N}=2$ supersymmetric QCD," hep-th/9806015; Nucl. Phys.
{\bf B 535} (1998) 512.}

\lref\kub{T. Kubota and N. Yokoi, ``RG flow near the superconformal
points in ${\cal N}=2$ supersymmetric gauge theories," hep-th/9712054;
Prog. Theor. Phys. {\bf 100} (1998) 423.}

\lref\hr{M. Mari\~no and G. Moore, ``The Donaldson-Witten function for
gauge groups of rank larger than one," hep-th/9802185;
Prog.Theor.Phys. {\bf 100} (1998) 423.}

\lref\bpv{W. Barth, C. Peters and A. Van de Ven, {\it Compact complex
surfaces}, Springer, 1984.}

\lref\beau{A. Beauville, {\it Complex algebraic surfaces}, Cambridge
University Press, 1996.}

\lref\gompfnc{R.E. Gompf, ``A new construction
of symplectic manifolds," Ann. of Math. {\bf 142} (1995) 527.}

\lref\taubes{C.H. Taubes, ``The Seiberg-Witten
invariants and symplectic forms," Math. Res. Letters
{\bf 1} (1994) 809; ``SW $\Rightarrow$ Gr: From the
Seiberg-Witten equations to pseudo-holomorphic curves,"
J. Amer. Math. Soc. {\bf 9} (1996) 845.}

\lref\ko{D. Kotschik, ``The Seiberg-Witten invariants of
symplectic four-manifolds [after C.H. Taubes],"
S\'eminaire Bourbaki, 48\`eme ann\'ee, 1995-6, n. 812. }

\lref\szabo{Z. Szab\'o, ``Simply-connected irreducible four-manifolds with
no symplectic structures," Invent. Math. {\bf 132} (1998) 457. }

\lref\fsknot{R. Fintushel and R.J. Stern, ``Knots, links, and
four-manifolds," dg-ga/9612004.}

\lref\fsrat{R. Fintushel and R.J. Stern, ``Rational blowdowns of smooth
four-manifolds," alg-geom/9505018, J. Diff. Geom. {\bf 46} (1997) 181.}

\lref\brussee{R. Brussee, ``The canonical class and the $C^{\infty}$
properties of K\"ahler surfaces,'' alg-geom/9503004. }

\lref\kmone{P.B. Kronheimer and T.R. Mrowka, ``Embedded
surfaces and the structure of Donaldson polynomials," J. Diff. Geom. {\bf
41} (1995) 573. }

\lref\fsconst{R. Fintushel and R.J. Stern, ``Constructions of
smooth four-manifolds," Doc. Math., extra volume ICM 1998, p. 443.}

\lref\kmad{P.B. Kronheimer and T.R. Mrowka,
``The genus of embedded surfaces in the projective plane," Math. Res. Lett.
{\bf 1} (1994) 797.}

\lref\jpark{J. Park, ``The geography of irreducible four-manifolds,"
Proc. Amer. Math. Soc. {\bf 126} (1998) 2493.}

\lref\mms{J.W. Morgan, T.R. Mrowka, and Z. Szab\'o, ``Product formulas along
$T^3$ for Seiberg-Witten invariants," Math. Res. Lett. {\bf 4} (1997) 915.}

\lref\mszabo{J.W. Morgan and Z. Szab\'o, ``Embedded
genus-$2$ surfaces in four-manifolds,"
Duke Math. Journal {\bf 89} (1997) 577.}

\lref\fsonly{R. Fintushel and R.J. Stern, ``Nonsymplectic
four-manifolds with one basic class,"  preprint.}

\lref\hh{F. Hirzebruch and H. Hopf, ``Felder von
Fl\"achenelementen in $4$-dimensionalen Mannigfaltigkeiten",  Math. Ann.
{\bf 136} (1958) 156.}

\lref\lloz{J.M.F. Labastida and C. Lozano, ``Duality
symmetry in twisted ${\cal N}=4$ supersymmetric gauge theories in four
dimensions," hep-th/9806032.}

\lref\gr{M. Mari\~no and G. Moore, ``The Donaldson-Witten
function for gauge groups of rank larger than one," hep-th/9802185.}

\lref\na{J.M.F. Labastida and M. Mari\~no, ``Non-abelian monopoles on
four-manifolds,"
hep-th/9504010; Nucl. Phys. {\bf B 448} (1995) 373. ``Polynomial invariants for
$SU(2)$ monopoles,"
hep-th/9507140; Nucl. Phys. {\bf B 456} (1995) 633.}

\lref\tqcd{S. Hyun, J. Park and J.S. Park, ``Topological QCD,"
hep-th/9503201; Nucl. Phys. {\bf B 453} (1995) 199.}

\lref\afre{D. Anselmi and P. Fr\'e, ``Topological sigma
model in four-dimensions and triholomorphic maps," hep-th/9306080; Nucl.
Phys. {\bf B 416}
(1994) 255. ``Gauge hyperinstantons and monopole equations,"
hep-th/9411205; Phys. Lett. {\bf B 347}
(1995) 247.}

\lref\tmfour{M. Alvarez and J.M.F. Labastida, ``Topological
matter in four dimensions," hep-th/9404115; Nucl. Phys. {\bf B 437} (1995)
 356.}

\lref\oko{C. Okonek and A. Teleman, ``Recent developments in
Seiberg-Witten theory and complex geometry," alg-geom/9612015, and
references therein.}

\lref\pt{V. Pidstrigach and A. Tyurin, ``Localisation of the Donaldson
invariants along Seiberg-Witten basic classes," dg-ga/9507004.}

\lref\fele{P.M.N. Feehan and  T.G. Leness, ``$PU(2)$ monopoles and relations
between four-manifold invariants," dg-ga/9709022; ``$PU(2)$
monopoles I: Regularity, Uhlenbeck compactness, and
transversality," dg-ga/9710032; ``$PU(2)$ monopoles II:
Highest-level singularities and relations between four-manifold
invariants," dg-ga/9712005.}

\lref\eqc{J.M.F. Labastida and M. Mari\~no, ``Twisted ${\cal N}=2$
supersymmetry
with central charge and equivariant cohomology," hep-th/9603169; Comm.
Math. Phys.
{\bf 185} (1997) 323.}

\lref\lalo{J.M.F. Labastida and C. Lozano, ``Mass perturbations in twisted ${
\CN}=4$ supersymmetric gauge theories," hep-th/9711132; Nucl. Phys. {\bf B
518} (1998)
37.}

\lref\dps{R. Dijkgraaf, J.S. Park and B.J. Schroers, ``${\cal N}=4$
supersymmetric
Yang-Mills theory on a K\"ahler surface," hep-th/9801066.}

\lref\GrHa{P.~ Griffiths and J.~ Harris, {\it Principles of
Algebraic Geometry},  J.Wiley and Sons, 1978. }

\lref\mmth{M. Mari\~no, ``The geometry of supersymmetric
gauge theories in four dimensions," hep-th/9701128.}

\lref\letter{M. Mari\~no, G. Moore and G. Peradze, ``Four-manifold
geography and superconformal symmetry," math.DG/9812042.}


\Title{\vbox{\baselineskip12pt
\hbox{YCTP-P26-98 }
\hbox{hep-th/9812055}
}} {\vbox{\centerline{Superconformal Invariance}
\centerline{}
\centerline{ and }
\centerline{}
\centerline{ The Geography of Four-Manifolds }}
}
\centerline{Marcos Mari\~no, Gregory Moore, and Grigor Peradze}

\bigskip
{\vbox{\centerline{\sl Department of Physics, Yale University}
\vskip2pt
\centerline{\sl New Haven, CT 06520, USA}}

\centerline{ \it marcos.marino@yale.edu }
\centerline{ \it moore@castalia.physics.yale.edu }
\centerline{ \it grigor.peradze@yale.edu }

\bigskip
\bigskip
\noindent
The correlation functions of supersymmetric gauge theories on a
four-manifold $X$ can sometimes be expressed in terms of
topological invariants of $X$. We show how the existence of
superconformal fixed points in the gauge theory can provide
nontrivial information about four-manifold topology. In particular,
in the example of gauge group $SU(2)$ with one doublet
hypermultiplet, we derive a theorem relating classical topological
invariants such as the Euler character and signature to sum rules
for Seiberg-Witten invariants. A short account of this paper can be
found in \letter.

\Date{Dec. 6, 1998}


\newsec{Introduction}

Expeditions in the current Age of Exploration of supersymmetric
quantum field theory have recovered a number of impressive
trophies. Chief amongst these has been the delivery of boatloads of
new conformal and superconformal fixed points in four (and even
higher) dimensions. At the same time, explorations of a rather
different nature into the application of supersymmetric quantum
field theory to the differential topology of   four-manifolds have
brought to light another collection of trophies: the Seiberg-Witten
invariants and the exact formulae for the Donaldson-Witten path
integral in $d=4, \CN=2$ topologically twisted gauge theories. In
view of this, one might hope that the new superconformal fixed
points will be a source of further insights into the topology of
four-manifolds. Conversely, advances in 4-manifold theory might
prove to be very useful in understanding aspects of superconformal
fixed points.

The present paper takes a small step in the program of combining
superconformal invariance with four-dimensional topological field
theory. We use the behavior of the Donaldson-Witten function at
superconformal points to prove an interesting and nontrivial
property of the topology of a compact, oriented, four-manifold $X$.
Loosely stated, our main result is the following. Let $\chi,\sigma$
be the Euler character and signature of $X$. Then, {\it either} $7
\chi + 11 \sigma \geq -12$, {\it or} the Seiberg-Witten invariants
of $X$ satisfy a collection of sum rules. The message is
summarized in Fig. 2 of section 6.2 below.
A more precise  statement
of the result can be found in the theorems in sections 6.1 and 6.3
below. The sum rules that we have found give very strong
constraints on the structure of the basic classes and the values of
the Seiberg-Witten invariants, and they show that the line $7 \chi
+ 11 \sigma = -12$ plays an important role in four-manifold
topology.

The organization of the paper is as follows.

 The evaluation of the
Donaldson-Witten function for $SU(2)$ gauge theories with $N_f=1,2,3,4$
fundamental hypermultiplets  was  carried out in \mw, but  the overall
normalization of the path
integral, which is   a function of the hypermultiplet masses and the quantum
scale $\Lambda_{N_f}$ was left undetermined.
We fill in this gap in section two  using the
standard tools of dimensional analysis, holomorphy, RG flow, and
anomalies. In the present paper we will be concentrating on the
superconformal fixed point of the theory with $N_f=1$ studied in
\apsw. In section three we summarize a few expansions in the
scaling variable $z$ near this fixed point. These expansions are
needed in the subsequent technical development. In section four we
combine the previous results to examine in detail the analytic
structure of the Donaldson-Witten function $Z_{DW}(z)$ as a
function of the scaling variable $z$. Our key physical observation
is that the function $Z_{DW}(z)$ must be {\it regular} at $z=0$.
Given the structure of $Z_{DW}(z)$ as a function of topological
invariants, this is a nontrivial fact. In section five we review for
the benefit of the reader (and the authors) some basic facts in the
 problem of the ``geography of four-manifolds.''
Loosely put, this is the problem of mapping out which regions in
the $\chi,\sigma$ plane are inhabited by four-manifolds and which
are not. As we shall see, there are definitely regions of {\it
terra incognita}. In section six we state our main result on
geography following from the analytic structure of $Z_{DW}(z)$. As
stated above we show that either $7 \chi + 11 \sigma \geq -12$,
{\it or} the Seiberg-Witten invariants of $X$ satisfy a collection
of sum rules. The precise statement of the sum rules is,
unfortunately, rather intricate. However, we introduce a notion of
a four-manifold of {\it superconformal simple type}. Roughly
speaking, these are manifolds which satisfy our sum rules in the
most natural way. As another byproduct of our results, we show that
manifolds with only one basic class have to satisfy $7 \chi + 11
\sigma \geq -12$. In section seven, we give
what we hope is compelling evidence for our
results: we study in detail complex surfaces, as well as many
important constructions of four-manifolds, and we show that all
four-manifolds of $b_2^+>1$ of simple type (of which we are aware)
are of superconformal simple type. In section eight, we discuss
upper and lower bounds on the number of basic classes for manifolds of
superconformal simple type. In particular, we
find a lower bound which generalizes the Noether inequality.
Finally, in section nine we state
our conclusions and some possible applications of our work.

\newsec{Remarks on the  $\alpha$ and $\beta $ functions }

In this somewhat technical
section we discuss the overall normalization of
the Donaldson-Witten partition function. For the sake
of brevity we will assume (in this section only)
some familiarity with the
results and notation of section 11 of \mw, which
discusses the $u$-plane integral for  gauge group
$SU(2)$ or $SO(3)$ in theories with matter
hypermultiplets.

When the matter hypermultiplets are in
the fundamental representation, the usual twisting procedure is
consistent only if the second Stiefel-Whitney class of the gauge
bundle equals that of the four-manifold, {\it i.e.} if $w_2(E)= w_2
(X)$. One can in fact consider more general twisting procedures (by
twisting for instance the baryon number of the hypermultiplet
\baryon), and this would allow us to study more general situations.
In this paper we will study only the usual twist, although the
generalization might be interesting. The $u$-plane integral for the
theories with matter hypermultiplets on a simply-connected
four-manifold has the form \mw:
\eqn\onea{Z_u(p,S;m_i,\Lambda) =
\int_{C}
{du  d\bar u \over  y^{1/2}}
\mu(\tau) e^{2 p u + S^2 \hat T(u)} \Psi,}
where the measure  $\mu (\tau)$ is given by
\eqn\oneb{
\mu(\tau)   = -{{\sqrt 2} \over 2}
 \alpha^\chi \beta^\sigma  { d \bar \tau \over  d \bar u}
\bigl({d a \over d u} \bigr)^
{ 1- \half  \chi  }
\Delta^{ \sigma /8}.
} In this equation, $\Delta$ is the discriminant of the
corresponding Seiberg-Witten curve, and is a polynomial in $u$,
$m_i$, $i=1, \dots, N_f$, and $\Lambda_{N_f}$ for the
asymptotically free theories with $N_f<4$.
The definitions of the other terms in \onea\ can
be found in \mw\ and will not be needed here.

The measure \oneb\
involves functions $\alpha, \beta $, which depend in principle on
$m_i, \Lambda_{N_f}$. \foot{ In the theory with $N_f=4$ doublet
hypermultiplets, as well as in the mass-deformed ${\cal N}=4$
theory, the $\alpha, \beta$ functions depend in principle on the
masses $m_i$ and the microscopic coupling constant $\tau_0$. The
mass-deformed ${\cal N}=4$ theory has been studied in detail from
the point of view of the $u$-plane integral in \lloz, where a
proposal for the $\alpha, \beta$ functions has been made for this
theory. In this paper, we will only consider the asymptotically
free theories, $N_f \leq 3$.} These functions can be obtained in the $N_f=0$
theory by comparing the results obtained via the $u$-plane integral
to the mathematical results in Donaldson theory, as   in \mw.
Our aim here is to be more precise  about the functions
$\alpha,\beta$. In
particular we will be interested in their mass dependence. In order
to do this, it is important to recall the physical origin of the
$\alpha, \beta$ functions in more detail \witteni\mw.

When the
twisted ${\cal N}=2$ theories are considered in a gravitational
background, there are extra couplings involving the curvature
tensor that can be generated in the effective action on the
$u$-plane. The requirement of topological invariance tells us that
the only possible allowed terms are in fact the signature and the
Euler  character densities. These densities have dimension $4$, and
will couple to holomorphic ``functions" in $u, m_i, \Lambda_{N_f}$,
therefore the effective action will contain the extra couplings
\eqn\densities{
( \log A (u, m_i, \Lambda_{N_f})) \chi + (\log B (u, m_i,
\Lambda_{N_f})) \sigma.} Notice that $A$, $B$ are dimensionless.
The $u$-dependence of these functions was determined in \mw, where
it was found that
\eqn\abs{
\eqalign{
A( u, m_i, \Lambda_{N_f}) &= \alpha (m_i, \Lambda_{N_f}) \biggl( {
du \over da} \biggr)^{1/2}, \cr
 B( u, m_i, \Lambda_{N_f}) &= \beta (m_i, \Lambda_{N_f}) \Delta^{1 / 8}. \cr}}
This explains the presence of these terms in the measure of the
$u$-plane integral, in \oneb.

To find the structure of the functions $\alpha$, $\beta$,
we have
to take into account some physical criteria that are similar to
those used in the analysis of superpotentials in ${\cal N}=1$
supersymmetric gauge theories (see \is\ for a review) and in the
derivation of the Seiberg-Witten curves in ${\cal N}=2$
supersymmetric gauge theories \swi\swii.

1. Holomorphy: $\alpha, \beta $ are local holomorphic functions of
$m_i$, $i=1, \dots, N_f$ and $\Lambda_{N_f}$.

2. Dimensional analysis: $Z_u$ is dimensionless.

3. RG-flow: in the double scaling limit where one of the
hypermultiplet masses goes to infinity in such a way that $m_{N_f}
\Lambda_{N_f}^{4-N_f}\equiv \Lambda_{N_f-1}^{4-(N_f-1)}$ is fixed,
we expect the general behavior,
\eqn\rg{
Z_u^{N_f} \rightarrow \biggl( {m_{N_f} \over \Lambda_{N_f}}
\biggr)^{\theta_{N_f}} Z_u^{N_f-1},} where $\theta_{N_f}$ is some
exponent that can depend on the number of flavors and on the
topological invariants $\chi, \sigma$.

The constraint of dimensional analysis on
 $\alpha, \beta$ is easily solved. We have canonical
dimensions for all the quantities involved in the Donaldson-Witten
generating function:
\eqn\onec{
\eqalign{
  [u] &= 2, \,\,\,\,\,\,\,\,\  [\tau ] =0,
\,\,\,\,\,\,\,\,\  [a] =1,  \cr
[\Psi]& =0 , \,\,\,\,\,\,\,\,\ [\Delta ] = 12. \cr } } The
dimension of $\Delta $ follows from the structure of the
Seiberg-Witten curve, where $[x]=2,[y]=3$. It follows from \abs\
that
\eqn\oned{
 [ \alpha ] = -1/2, \,\,\,\,\,\,  [\beta ] =-3/2.
}
Notice
that, for $Z_u$ to be dimensionless, we have to include an overall
factor $1/(c_{N_f} \Lambda_{N_f})$ in the $u$-plane measure, where
$c_{N_f}$ is a constant.

Imposing the other constraints is more involved
and depends on $N_f$.
We will give two arguments that $\alpha,\beta$ are independent
of the masses and given by
$\alpha = \alpha_0/
\Lambda_{N_f}^{1/2}$, and $\beta=\beta_0/\Lambda_{N_f}^{3/2}$,
where $\alpha_0, \beta_0$ are constants.
The second argument only applies to the case
$N_f=1$, which is the case of primary interest
here.

First of all, one can argue on general grounds that the
functions $\alpha, \beta$ must be independent of the masses. If
they had such a dependence, since they are locally holomorphic functions of
the masses, they would have singularities at some special values
$m^{*}_i$, and for {\it any} value of $u$. Therefore, the functions
$A( u, m_i, \Lambda_{N_f})$ and $B( u, m_i, \Lambda_{N_f})$
appearing in the $u$-plane integral would then have extra
singularities along complex codimension one varieties of the form
$(m^{*}_i, u)$, for any $u \in \IC$. But there is no physical
reason to expect such behavior on manifolds of
$b_2^+(X)=1$. The only points in the $(m_i, u)$
space where we can have zeroes or poles for these functions are at
the singularities in the $u$-plane, which have the form $(m_i,
u_*(m_i))$ (where the $u_* (m_i)$ are the zeroes of the
discriminant $\Delta (u, m_i, \Lambda_{N_f})$.) One could think
that there are in fact special values of the masses, namely the
critical values giving superconformal fixed points, where some
singular behavior can show up. But, again, the superconformal
points are a special class of singularities in the $u$-plane, and
they only occur for special values of $u$. Moreover, if we consider
the contributions of the SW singularities (as we will do in this
paper), the $\alpha$ and $\beta$ functions will be overall factors
for the contribution of all the singularities. If they had zeroes
or poles at the values of the $m_i$ associated to the
superconformal points, for example, we would find a singular
behavior even in the contributions of the singularities not
involved in a nonlocal collision. This is clearly unreasonable
physically. \foot{Notice that the same arguments rule out any mass
dependence in the overall factor that we introduced in the measure
for dimensional reasons.}

The argument we have just given is subject to a
possible subtlety for the theories $N_f>1$ related
to the appearence of noncompact Higgs branches.
Thus, while we regard the argument as reasonable,
we also present a second argument which applies
in the case $N_f=1$ and leads to the same result.

The  second argument proceeds by using
 another physical input which is crucial in the analysis of
the functions $A$, $B$   explained in \witteni\mw.  These
functions have to satisfy the general constraints listed above, but
they also have to reproduce the anomaly associated to the fields
that have been integrated out.  In fact, we can understand the
terms in \abs\ as anomaly functionals in the effective theory that
reproduce the anomaly in the $R$-current due to a background
gravitational field. This physical input was enough to find the
$u$-dependence of $A$, $B$ in \witteni\mw. As we will see, a slight
modification of this analysis also fixes the dependence of $\alpha,
\beta$ on $m$, $\Lambda_{N_f=1}$.

We want to analyze the terms $A$, $B$ using the general
constraints above as well as the anomaly condition. Since the
hypermultiplet mass explicitly breaks the $U(1)_R$ symmetry, in
order to have the classical $R$-symmetry we use Seiberg's trick and
give $R$-charge two to the mass. Therefore, under a $U(1)_R$
rotation, we have that:
\eqn\uonerot{
u \rightarrow {\rm e}^{4i \phi_R} u, \,\,\,\,\,\,\,\  m
\rightarrow {\rm e}^{2i \phi_R} m.} The scale $\Lambda_{N_f=1}$
doesn't rotate under $U(1)_R$. Let's now consider the anomaly
analysis of $A$. By dimensional analysis, we know that $1/\alpha^2$
is a homogeneous function of degree one in $\Lambda_{N_f=1}$, $m$,
that we will denote by $P(m, \Lambda_{N_f=1})$. It will be enough
to analyze the behavior at the semiclassical region, where
$(du/da)^{1/2} \sim u^{1/4}$. The term
\eqn\semi{
\log \biggl( {u^{1/4} \over P(m, \Lambda_{N_f=1}) }\biggr) \chi}
has to reproduce the part of the anomaly proportional to $\chi$,
and corresponding to the massive $W$ bosons that have been
integrated out. This means that the function of $u$,
$\Lambda_{N_f=1}$ and $m$ in the argument of the logarithm must
have $R$-charge $1$ \witteni\mw. As the mass has $R$-charge two now, the
function $P (m, \Lambda_{N_f=1})$ has to be  independent of $m$. This means
that $\alpha$
is independent of the mass, and is given by $\alpha = \alpha_0/
\Lambda_{N_f=1}^{1/2}$, where $\alpha_0$ is a constant.

A similar analysis can be done for $B$, where the anomaly in the
semiclassical region corresponds now to the massive components of
the hypermultiplet and depends on $\sigma$. Again, to reproduce the
anomaly $\beta$ must be independent of $m$, and we obtain
$\beta=\beta_0/\Lambda_{N_f=1}^{3/2}$, $\beta_0$ a constant.

Using the expressions for $\alpha$,
$\beta$ that we have derived, as well as the RG relation
$\Lambda_{N_f}^{4-N_f} m_{N_f}=\Lambda_{N_f-1}^{5 -N_f}$, we find
that in fact the $u$-plane integral changes as in $\rg$, with
\eqn\thetanf{
\theta_{N_f} = { 3 + \sigma \over 5 - N_f},}
where we have used that $\chi+ \sigma =4$ for simply-connected
four-manifolds with $b^+_2=1$.

Finally, we remark that the $\alpha,\beta$ functions can in
principle be derived from a weak-coupling one-loop calculation at
large values of $u$. It should also be possible to derive them from
field theory limits of string gravitational corrections such as the
quantum correction $F_1 \tr R \wedge R$ in $\CN=2, d=4$
compactifications of string theory. The string theory approach to
deriving the $\alpha,\beta$ functions raises many interesting further
issues which are beyond the scope of this paper.

\newsec{Some properties of the superconformal points}

\subsec{Superconformal divisors in the $SU(2)$ theories with matter}
At some special divisors in the moduli space of ${\cal N}=2$
supersymmetric gauge theories, the low-energy theory is a
non-trivial superconformal field theory. The first example of such
a point was found in \ad\ in pure $SU(3)$ Yang-Mills, and
subsequently many other superconformal divisors were found in the
$SU(2)$ theories with matter \apsw, as well as in higher rank
theories with matter \eguchisc. These superconformal divisors are
characterized by the fact that at them, two or more mutually
non-local BPS states become massless simultaneously.

In the $SU(2)$ theories with $N_f$ massive hypermultiplets, the
superconformal points appear as follows: for generic values of the
masses, there are $2 + N_f$ points in the $u$-plane with massless
BPS states. For special values of the masses, these singularities
can collide. There are two possibilities for these collisions. The
first possibility is that the $k$ states that come together are
mutually local, and the low-energy theory will be ${\cal N}=2$ QED
with $k$ hypermultiplets. This happens, for instance, in the
massless case, and more generally along the lines where $m_i={\pm }
m_j$. The second possibility is to tune the values of the
hypermultiplet masses, in such a way that the singularities that
collide are mutually non-local. This will give nontrivial
superconformal points. When two singularities collide to give a
superconformal point, one of them will correspond to $k$ mutually
local massless states, while the other one will be associated to a
massless state which is mutually non-local with respect to the $k$
states in the other singularity. For this reason, these points are
called $(k,1)$ points \apsw. The $SU(2)$ theories with $N_f < 4$
hypermultiplets will have superconformal points of type $(k,1)$ for
$k=1, \dots, N_f$. The $N_f=4$ theory has no additional collisions,
and it will have $(k,1)$ points with $k=1,2,3$. Some of these
superconformal points are the following:

a) In the $N_f=1$ theory, there are $(1,1)$ superconformal points
when the mass of the hypermultiplet satisfies $m_*^3 = (3
\Lambda_1/4)^3$, and these points are located at $u_*^3= (3
\Lambda_1^2/4)^3$.

b) In the $N_f=2$ theory, there are $(2,1)$ points that occur when
$m_1=\pm m_2$ and $m_1^2=\Lambda^2_2/4$.

c) Finally, in the $N_f=3$ theory, a $(3,1)$ point appears when
$m_1=m_2=m_3=\Lambda_3/8$.

The critical exponents and scaling dimensions of the operators in
these superconformal theories can be obtained by looking at the
structure of the Seiberg-Witten curves near the colliding
singularities \apsw\ or by using the expansions of the
Seiberg-Witten periods in the critical theory \bilalfer\kub, and
are in fact completely determined by the value of $k$ .

\subsec{Expansions around the superconformal point}

In this paper, we will be interested in the simplest superconformal
point arising in the $SU(2)$ theories with matter: the
superconformal point of type $(1,1)$ in the $N_f=1$ theory. For
simplicity, we will only consider the $(1,1)$ point in the $m$-plane,
occurring at $m_*= 3 \Lambda_1/4$ (the other
$(1,1)$ points are obtained by $\IZ_3$ symmetry on the complex
$m$-plane). We want to study the behavior of the theory as $m$
approaches the critical value $m_*$. As we will see, the quantities
appearing in the topological correlation functions are natural
quantities associated to the elliptic curve describing the
low-energy theory, and when $b_2^+>1$ and the manifold is of simple
type, they involve the evaluation of these quantities at the
singularities in the $u$-plane. When the theory is nearly critical
and $m =m_* +z$, the $u$-coordinates of the two singularities that
collide at the superconformal point (that will be denoted by
$u_{\pm}$) can be expanded in terms of the parameter $z$.
Therefore, all the quantities evaluated at $u_{\pm}$ will have an
expansion in $z$ as well. We will present here explicitly the first
few terms of these expansions, since we will need them in the next
section.

We have to evaluate some quantities associated to the
Seiberg-Witten curve at the $u$-plane singularities. To do this, we
follow the procedure in section 11 of \mw\ and write the SW curves
\eqn\swcurvegnrl{
y^2=x^3+a_2 x^2+a_4 x +a_6 } in the form
\eqn\red{
y^2=x^3- { c_4 \over 48} x - { c_6 \over 864}, } where
\eqn\coeffs{
\eqalign{
c_4 & = 16 (a_2^2 -3 a_4), \cr
 c_6 & = -64 a_2^3 + 288 a_2 a_4 -864 a_6.}}
The value of the period at the singularity $u=u_*$ is given by
\eqn\period{
\bigl( { da \over du} \bigr) ^2 _*= { c_4  (u_*) \over 2 c_6 (u_*)}.}
Another quantity that enters in the Donaldson-Witten function for
manifolds of simple type is the following. Consider the effective
coupling $\tau$ in the duality frame appropriate to the singularity
$u=u_*$ (in particular, $\tau \rightarrow i \infty$ as $u
\rightarrow u_*$). We define $\kappa$ as $\kappa = (d u / d q)$,
where $q={\rm e}^{2 \pi i \tau}$. The value of this quantity at the
singularity can be expressed in terms of  \coeffs\ and the
discriminant of the curve as
\eqn\kap{
\kappa_*= {c_4^3 (u_*) \over \Delta ' (u_*) }. }

We will now consider the behavior of these quantities (evaluated at
the $u$-plane singularities) for the $N_f=1$ theory with a massive
hypermultiplet, when the value of the mass is near the critical
value $m_*= 3 \Lambda_1/4$. The Seiberg-Witten curve  in this
case is \swii:
\eqn\thecurve{
y^2 = x^2(x-u) + {1 \over 4} m \Lambda_1^3 x - {1 \over 64}
\Lambda_1^6,} with discriminant
\eqn\discone{
\Delta_1 (u,m, \Lambda_1) = {\Lambda_ 1^6 \over 16} \biggl[ -u^3 + m^2 u^2 +
{ 9 \over 8} \Lambda_1^3 mu - \Lambda_1^3 m^3 - { 27 \over 256} \Lambda_1^6
\biggr].}
One finds
\eqn\oneco{
\eqalign{
c_4(u,m,\Lambda_1) &= 16 u^2 -12 \Lambda_1^3 m,\cr
 c_6(u,m,\Lambda_1) &=64 u^3 -72 \Lambda_1^3 mu + {27 \over 2}
\Lambda_1^6.\cr}}
For the critical value of the mass $m=m_*= 3 \Lambda_1/4$, two
singularities on the $u$-plane will collide at $u_*=3
\Lambda_1^2/4$. If we write
\eqn\scdc{
m=m_*+z, \,\,\,\,\,\  u =u_*+\Lambda_1 z+ \delta u, } and we
introduce the shifted variable $ x=u/3 + \tilde x$, the
Seiberg-Witten curve \thecurve\ becomes, at leading order,
\eqn\cuspidal{
y^2= \tilde x^3 - {\Lambda_1^3 \over 4} z \tilde x -{ \Lambda_1^4
\over 16}\delta u,} which is a deformation of the cuspidal cubic
$y^2 = \tilde x^3$. The variables $z$ and $\delta u$ correspond to
operators in the conformal field theory at the $(1,1)$ singularity.
The scaling dimensions of these operators can be deduced from
\cuspidal\ after taking into account that $a\sim (\delta  u/y) d\tilde x$
has dimension $1$, and one finds that $z$ has dimension $4/5$,
while $\delta u$ has dimension $6/5$.

If $u_{\pm}=u_*+\Lambda_1 z+ \delta u_{\pm}$ denotes the position
of the two colliding singularities, the deformation parameter
$\delta u_{\pm} $ will depend on $z$ as well, and of course $\delta
u_{\pm} \rightarrow 0$ when $z \rightarrow 0$. The dependence of
$\delta u_{\pm}$ on $z$ can be obtained as a power series expansion
by looking at the zeros of the discriminant \discone, $\Delta_1
(u,m, \Lambda_1)=0$ in terms of the variables \scdc. The most
appropriate normalization for $\Lambda_1$, for our purposes, is $4 {\sqrt 3}
\Lambda^{3/2}_1=1$ (with this normalization, the leading term
of $(du/da)^2$ is $z^{1/2}$.) One then finds,
\eqn\scdd{
\delta u_{\pm}  =
\pm
\biggl( {16 \over 243} \biggr)^{1/3} z^{3/2}+
{4 \over 9 } z^2 + {\cal O}( z ^{5/2} ). } Notice that
\eqn\interchan{
\delta u_- (z^{1/2}) = \delta u_+ (-z^{1/2}).}
This is due to the fact that, for $z<0$, one has $\pm z^{1/2}= \pm
i |z|^{1/2}$. But in this case the roots $u_{\pm}$ of the equation
$\Delta_1 (m,u, \Lambda_1)=0$ must be related by complex
conjugation, therefore we must have \interchan. Also notice that
the leading term of \scdd\ follows from the structure of the curve
near the $(1,1)$ singularity given in \cuspidal. Using the
expansion \scdd\ and the explicit expressions \oneco\period\kap, we
have the following expansions of $(du/da)^2$ and $\kappa$ at the
singularities $u_{\pm}$ for $m=m_*+ z $:
\eqn\scddone{
\eqalign{
\bigl({du \over{da}}\bigr)_{\pm}^2 &=
\pm z^{1/2} \biggl\{ 1  \pm \bigl( { 4 \over 3} \bigr)^{4/3} z^{1/2}
+ {\cal O} (z) \biggr\} \cr
\kappa_{\pm} &=
\mp 2^{31/3} \cdot 3^{4/3}  z^{3/2}
\biggl\{ 1  \pm \bigl( { 2048 \over 3} \bigr)^{1/3} z^{1/2}
+ {\cal O} (z) \biggr\} .\cr
} } Notice that, due to \interchan, one has the following property:
\eqn\prop{
\bigl({du \over{da}}\bigr)_{-}^2 (z^{1/2})
= \bigl({du \over{da}}\bigr)_{+}^2 (-z^{1/2}),}
and a similar equation for $\kappa_{\pm}$. This will be important
when we consider the analytic properties of the correlation
functions as a function of $z$. Another important  remark is that
the leading powers of $z$ in the expansion of $\delta u$,
$(du/da)^2$ and $\kappa$ are determined by the anomalous scaling
weights of the operators near the $(1,1)$ superconformal point.

\newsec{Correlation functions near the superconformal point}

In this section, we will study the behavior of the Donaldson-Witten
function for the $SU(2)$ theory with one massive hypermultiplet,
and for manifolds with $b_2^+>1$ and of simple type. The general
answer for the generating function in the theories with matter was
obtained in \mw, and we briefly review it in 4.1. Using this
expression, we will study the behavior of the generating function
near the superconformal point. In 4.2 we extract from the
generating function a Laurent expansion in $z$, that we denote by
$F(z)$, and we analyze some of its properties. In 4.3 we argue on
physical grounds that this Laurent expansion is in fact a Taylor
expansion (equivalently, that the generating function is regular at
the superconformal point). This will be the main result of our
paper. In the rest of the sections, we will develop the
consequences of this fact for the geography of four-manifolds and
the structure of the SW invariants.

\subsec{The Donaldson-Witten function with massive hypermultiplets}

 The Donaldson-Witten function of  $\CN =2$, $SO(3)$ theories with matter
hypermultiplets can
be  written explicitly  in terms of the SW invariants, for
manifolds with $b_2^+>1$ and of simple type, but not necessarily
simply connected. The expression   for generic values of the masses
is:
\eqn\dwf{
\eqalign{
Z(m_i, \Lambda_{N_f}; p,S) & =2^{1 + { 3 \sigma + \chi \over 2}}
(-i)^{\chi_h} \biggl({ \pi^2 \beta^8 \over 2^8} \biggr)^{\sigma/8}
\biggl({ - \pi \alpha ^4 \over 2} \biggr)^{ \chi/4}
\sum_{j=1, \dots, 2+ N_f}  \kappa_j^{\chi_h}
\biggl( {da \over du} \biggr)_j ^{ -(\chi_h + \sigma)}\cr
& \cdot \sum_{\lambda} SW (\lambda) \exp \biggl[ 2 pu_j+ S^2 T_j -
 i \bigl( {du \over da} \bigr)_j (S, \lambda) \biggr]
{\rm e}^{ 2\pi i (\lambda_0^2 + \lambda \cdot \lambda_0)}. \cr}
}

The notation is the following:

\item{1.} The sum on $j$ is
 a sum over the $2 + N_f$ singularities at finite values on the $u$-plane.
The subindex $j$ in the different quantities means that they are
evaluated at the $j$-th singularity. The values of $da/du$ and
$\kappa$ at the singularities of the $u$-plane are given in
\period\ and \kap. The contact term at the singularity can be
written as \mw
\eqn\contact{
T_j=-{1 \over 24} \biggl( \bigl( {du \over da} \bigr)_j^2 -8 u_j
\biggr).}

\item{2.}  $\chi_h$ denotes
\eqn\thec{
\chi_h = {\chi + \sigma \over 4} = { 1- b_1 + b_2^+ \over  2} .}
Note in particular that we have not assumed $b_1=0$ and hence
$\chi_h$ can be negative. The notation for this particular
combination of $\chi$, $\sigma$ comes from the fact that, if the
four-manifold we are considering is a complex surface, then
$\chi_h$ is the holomorphic Euler characteristic.

\item{3.} The sum over $\lambda$ is a sum over
basic classes.  As explained in \mw, the theories with matter can
be considered on any smooth, compact, oriented four-manifold $X$ if
the non-abelian magnetic flux of the $SO(3)$ gauge bundle $E$
satisfies $w_2(E)=w_2 (X)$, where $w_2(E)$, $w_2(X)$ are the second
Stiefel-Whitney classes of the bundle $E$ and of the manifold $X$,
respectively. In the expression \dwf, we have to choose an integer
lifting of $w_2(X)$ (this lifting always exists, by a theorem of
Hirzebruch and Hopf \hh), which is denoted by  $c_0=2 \lambda_0$.
Notice that $\lambda \in \Gamma + {1 \over 2} w_2 (X)$, where
$\Gamma = H^2 (X; \IZ)$. Therefore, $x = 2 \lambda$ is an integral
cohomology class which is congruent to $w_2 (X)$ mod $2$, in other
words, $x$ is the first Chern class of the determinant line bundle
associated to a ${\rm Spin}^c$-structure. In particular, $x$ is
characteristic. The exponential involving $\lambda_0$ in \dwf\ can
be written then as
\eqn\phase{
(-1)^{ c_0 \cdot x + c_0^2 \over 2},} and since $x$ is
characteristic one can easily prove that the exponent is in fact an
integer. A change of the lifting $c_0 \rightarrow \tilde c_0$
multiplies the above generating function by the factor
\eqn\factor{
(-1)^{\bigl({\tilde c_0 - c_0 \over 2}\bigr)^2}.}

\bigskip
\noindent
{\bf Remarks}:

\item{1.} Equation  \dwf\
generalizes Witten's expression \monopole\ for pure Yang-Mills and
was obtained in section 11 of \mw. The extension of the $N_f=0$
Donaldson-Witten function to non-simply connected four-manifolds
was begun in
\mw\lns, and completed in \nsc.
The extension to $N_f>0$ is trivially obtained  following the
arguments in \nsc, where the $u$-plane integral and the SW
contributions were determined in the nonsimply connected case for
$N_f=0$. Somewhat later the same result for $N_f=0$ was stated,
less precisely, in \munnsc.

\item{2.} Notice that, if a manifold is of simple type and
has non-trivial SW invariants,
the dimension of the moduli space of solutions to
the SW monopole equations has to be zero,
and this implies that the index of the Dirac operator
coupled to the ${\rm Spin}^c$-structure
associated to $\lambda$ is:
\eqn\drcndx{
\half \lambda^2 - {\sigma \over  8 } = \chi_h
} where we have used  $(2 \lambda)^2 = 2 \chi + 3 \sigma$.
Therefore, if \dwf\ is not identically zero, $\chi_h$ must be an
integer \monopole.

\item{3.} For nongeneric masses (for instance, in the massless case)
and $N_f>1$, there are singularities where two or more mutually local
states become massless simultaneously. The analysis of the $u$-plane
integral in this situation has been done in \ky. Notice that at
these singularities, the monopole equations involve more than
one spinor field, giving rise to some generalized
Seiberg-Witten invariants. These invariants have not
been studied in detail, but we should expect a noncompact
moduli space of solutions, in correspondence with the
noncompact Higgs branch in the physical theory.

\item{4.} The UV theory with $N_f=1$ hypermultiplet
describes the nonabelian monopole theory. This theory
has been formulated from the point of view of topological
quantum field theory in \na\tqcd\tmfour\afre, and
from a mathematical point of view in \oko\pt\fele.
The mass terms for the hypermultiplet
have been interpreted in \eqc\lns\lalo\dps\ as an
equivariant extension of the Thom form with
respect to a $U(1)$ action on the moduli space of
nonabelian monopoles. For a detailed review of the different
aspects of the nonabelian monopole theory, see \mmth.

\subsec{The function $F(z)$}

We will now concentrate on the generating function \dwf\ for the
theory with one massive hypermultiplet. For generic values of the
mass, there are  three different singularities on the $u$-plane. At
each singularity, one of the periods of the Seiberg-Witten curve
goes to infinity (corresponding to the degeneration of one of the
cycles of the elliptic fiber over the $u$-plane), while the period
associated to the appropriate duality frame near the singularity
remains finite and different from zero. Likewise, $\kappa$
is finite and well-defined. Therefore, the expression in \dwf\ is
well-defined. A different situation arises when we approach a
superconformal point, $m \rightarrow m_*$. In this case, at the
colliding singularities, both periods go to infinity and \period\
diverges, as we have seen in the previous section. Notice that the
contribution to \dwf\ of the singularity which is not involved in
the collision is still perfectly regular as we approach the
critical value of the mass, since nothing special happens there. As
we showed in section 2, the $\alpha, \beta$ functions only depend
on the scale $\Lambda_1$, so they don't affect the behavior of the
generating functional as a function of the mass, and we will drop
them as well as the other constant factors in \dwf. Therefore,
to study the behavior at $z=0$ of the generating function, where
$z$ is defined in \scdc, we can
focus on the contribution of the two colliding singularities
$u_{\pm}$. This contribution is given by:
\eqn\twosing{
 \sum_{\pm}  \kappa_{\pm}^{\chi_h}
 \biggl[  \bigl( {du \over da} \bigr) ^2_{\pm}  \biggr] ^
{ {\chi_h + \sigma \over 2} } {\rm e}^{  2 p u_{\pm} + S^2 T_{\pm}
}\sum_{\lambda} SW (\lambda) {\rm e}^{- i \bigl( {du \over da}
\bigr)_{\pm} (S, \lambda)} {\rm e}^{ 2\pi i (\lambda_0^2 + \lambda
\cdot \lambda_0)}. }
We can now study the properties of this
function as a power series in $z$, focusing on the
issue of regularity at $z=0$.

First, we   express $\chi , \sigma$
in terms of more convenient
linear combinations:
\eqn\vari{
\eqalign{
\chi_h & = { \chi + \sigma \over  4}, \cr
c_1^2  & = 2 \chi + 3 \sigma.  \cr}
}
The quantity
$\chi_h$ was introduced in \thec.
When the four-manifold is a complex
surface (or, more generally, an almost complex manifold),
the
combination $c_1^2$
is the square of the first Chern class of the
holomorphic tangent bundle, but we will still use the above
notation for this particular combination of $\chi$ and $\sigma$ for
an arbitrary compact, oriented four-manifold. If the manifold $X$
is of simple type, then $x^2 =c_1^2$ for any basic class. Notice
that we can express $\chi$ and $\sigma$ in terms of $\chi_h$ and
$c_1^2$ as follows:
\eqn\reex{
\chi= 12 \chi_h -c_1^2, \,\,\,\,\,\,\ \sigma=c_1^2 -8 \chi_h.}
The reason that we introduce these combinations is to facilitate
comparison with the results on the geography of four-manifolds.

In
terms of $\chi_h$, $c_1^2$, the expansion of the factors that are
independent of $\lambda$ in \twosing\ reads as follows:
\eqn\leading{
\eqalign{
 &  \kappa_{+}^{\chi_h} \biggl[  \bigl( {du \over da} \bigr) ^2_{+}  \biggr] ^
{ {\chi_h + \sigma \over 2} } {\rm e}^{  2 p u_{+} + S^2 T_{+} }
\cr & \,\,\,\,\,\,\,\,\ = c^{\chi_h} {\rm e}^{u_{*} (2p+ S^2 /3)}
z^{ c_1^2 - \chi_h \over 4} \biggl\{ 1 +  \biggl( \bigl( { 4 \over
3} \bigr)^{4/3} \Bigl({ c_1^2 + 5 \chi_h \over 2}\Bigr)  -{S^2
\over 24 } \biggr) z^{1/2} + {\cal O} (z) \biggr\}, \cr &
\kappa_{-}^{\chi_h} \biggl[  \bigl( {du \over da} \bigr) ^2_{-}
\biggr] ^ { {\chi_h + \sigma \over 2} } {\rm e}^{  2 p u_{-} + S^2
T_{-} } \cr & \,\,\,\,\,\,\,\,\
=c^{\chi_h} {\rm e}^{u_{*} (2p+ S^2 /3)} {\rm e} ^{{\pi i \over 2}
(c_1^2-\chi_h)} z^{ c_1^2 - \chi_h \over 4}
\biggl\{ 1 -  \biggl( \bigl({ 4 \over 3} \bigr)^{4/3}
\Bigl( { c_1^2 + 5 \chi_h \over 2}\Bigr) -{S^2 \over 24 } \biggr) z^{1/2}
 + {\cal O} (z) \biggr\}, \cr}}
where $c= -(2^{31/3} \cdot 3^{4/3})$ is an overall constant in the
expansion of $\kappa_{\pm}$. The power of the leading term in $z$
in the expansions \leading\ is
\eqn\elsev{
{c_1^2 - \chi_h \over 4} = { 7 \chi + 11 \sigma \over 16}.} Notice
that the change of sign in $z^{1/2}$ introduces a relative phase
between the contributions at $u_{\pm}$. It is clear that, to study
the regularity of \twosing, we don't have to worry about the
overall $c^{\chi_h} \exp[u_{*} (2p+ S^2 /3)]$. We will define the
function $F(z)$ as
\eqn\functionf{
\eqalign{
F(z)& =\sum_{\pm}  (c^{-1}\kappa_{\pm})^{\chi_h} \biggl[  \bigl(
{du \over da} \bigr) ^2_{\pm}  \biggr] ^{ {\chi_h + \sigma \over 2}
} {\rm e}^{  2 p (u_{\pm}-u_*) + S^2 (T_{\pm}-u_*/3) } \cr
&\,\,\,\,\, \cdot \sum_{\lambda} SW (\lambda) {\rm e}^{- i \bigl(
{du \over da} \bigr)_{\pm} (S, \lambda)} {\rm e}^{ 2\pi i
(\lambda_0^2 + \lambda \cdot \lambda_0)}.\cr} }
The generating function of the $N_f=1$ theory will be regular at
the superconformal point if and only if $F(z)$ is regular at $z=0$.
In the rest of this section, we will focus on the properties of
$F(z)$.

First, we will find the structure of $F(z)$ as a power series. In
principle, as $F(z)$ contains the quantity $(du/da)_{\pm}$, the
expansion is in powers of $z^{1/4}$.
However, as we now show,
$F(z)$ in fact has an expansion in integral
powers of $z$. Consider the
$\lambda$-dependent piece of \functionf,   denoted by :
\eqn\swmame{
\CS\CW^{\pm} (z^{1/4})=\sum_{\lambda} SW (\lambda)
{\rm e}^{- i \bigl( {du \over da} \bigr)_{\pm} (S, \lambda)}
{\rm e}^{ 2\pi i (\lambda_0^2 + \lambda \cdot \lambda_0)}.} An
important property of these functions is that, if $\chi_h + \sigma$
is even (odd), they only contain even (odd) powers of
$(du/da)_{\pm}$. This is due to the fact that, if $\lambda$ is a
basic class, then $-\lambda$ is also a basic class, and \monopole
\eqn\rel{
SW(-\lambda) = (-1)^{\chi_h} SW (\lambda).} On the other hand,
changing $\lambda$ to $-\lambda$ in the phase in \swmame\
introduces a global factor
\eqn\glphase{
{\rm e}^{-4 \pi i \lambda_0 \cdot \lambda} = (-1) ^\sigma, } as one
can easily check using Wu's formula. If one decomposes $F(z)$
as the sum of the contributions at $u_{\pm}$, $F(z)=F_+ (z) + F_
{-}(z)$, it follows from the above analysis that  $F_{\pm}(z)$ only
contain even powers of $(du/da)_{\pm}$, respectively. Therefore, as
$\chi_h$ is an integer, the functions $F_{\pm}$ have a series
expansion in powers of $z^{1/2}$. Actually, more is true. As we
remarked in section 3.2, we have $F_-(z^{1/2}) = F_+ (-z^{1/2})$,
therefore $F(z)$ has in fact an expansion in integral powers of
$z$. We have proved the following

\bigskip
\noindent
{\bf Proposition 4.2.1}. The function $F(z)$ defined in \functionf\ has a
Laurent series expansion in integral powers of $z$ around $z=0$,
{\it i.e.}, there is no monodromy around $z=0$.

\bigskip
We finish this section by rewriting the expansion of $F(z)$ in a
way that will be useful in section 6.1.
Inserting the expansions \leading\ in $F(z)$, we obtain the
expression:
\eqn\expans{
\eqalign{
& F(z)= z^{ c_1^2 - \chi_h \over 4} \biggl\{ \CS\CW^+ (z^{1/4}) +
{\rm e} ^{{\pi i \over 2} (c_1^2-\chi_h)} \CS\CW^{-} (z^{1/4}) \cr
& + z^{1/2} \biggl( \bigl({ 4 \over 3} \bigr)^{4/3}\Bigl(  { c_1^2
+ 5 \chi_h \over 2}\Bigr)  -{S^2 \over 24 } \biggr) \bigl[ \CS\CW^+
(z^{1/4})- {\rm e} ^{{\pi i \over 2} (c_1^2-\chi_h)} \CS\CW^{-}
(z^{1/4}) \bigr]  + {\cal O} (z) \biggr\}. \cr}}
Notice that
the functions $\CS\CW^{\pm}
(z^{1/4})$ are themselves series expansions in $z^{1/4} $. Nevertheless,
we will
see, however, that this way of organizing the terms turns out to be
very useful. It is clear that the possible poles of this expression
are due to the power of $z$ in front of the expression, as all the
terms inside the curly brackets are regular at $z=0$. Also notice
that the terms that we haven't written explicitly involve linear
combination of $\CS \CW^{\pm}$ with coefficients that depend on
$p$, $S^2$, $\chi_h $ and $c_1^2$, and to write them we need the
explicit expansions of $(du/da)$ and $\kappa$ to arbitrary order.

\subsec{The main result: F is regular}

We now state the main result in this paper. We are studying the
theory with one massive quark hypermultiplet on a compact, oriented
manifold $X$ of simple type, with $b_2^+>1$. There are two sources
of divergence in the correlation functions of a topological quantum field
theory: one is due to the noncompactness of spacetime, and the
other is the noncompactness of a moduli space of vacua. In
our case, neither source of divergence is present: by assumption,
the spacetime manifold is compact. On the other hand, when
$b_2^+>1$, as it was proved in \mw, the only contributions to the
correlation functions come from a finite set of points in the
$u$-plane, and the moduli space of vacua associated to the
low-energy theory   -the moduli space of the SW monopole equations-
is also compact. We then have the following

\bigskip
\noindent
{\bf (Physical) Theorem 4.3.1}.  The function $F(z)$ is {\it analytic} at
$z=0$, {\it i.e.}, it has a Taylor series expansion around the
origin.
\bigskip

One of the key physical ingredients to guarantee the regularity of $F(z)$ is
the absence of Higgs branches in the moduli space, in the case
of $N_f=1$. These branches are noncompact
and could be a source of divergences in the path integral. For the
theories with $N_f \ge 2$, there are superconformal points of type
$(k,1)$, with $k>1$. In this case, the low-energy colliding singularity with
$k$ mutually local hypermultiplets will have a noncompact Higgs branch,
therefore
we can not guarantee the regularity of the generating function at the
superconformal
point. This noncompactness of the moduli space of vacua is the source of the
noncompactness of the moduli space of solutions to the monopole equations
with $k>1$ monopole fields. See, for examples, eq. (6.4) of
\lns\ or  equations   (3.26) and
(3.27) of \ky. When there is more than one
term on the right-hand-side of the $F^+ = \sum \bar M \Gamma M$ equation then
the moduli space is noncompact.

One could worry that, although
for $N_f=1$ the moduli space of vacua is
compact for any value of the mass arbitrarily
close to its critical value,
a Higgs branch might appear in the moduli space precisely when $m=m_*$. Indeed,
the phenomenon of
a   ``jump''  in the Higgs branch for special values
of parameters is a well-known
phenomenon in supersymmetric gauge theory and
string theory.  Indeed, when {\it mutually local}
singularities collide the Higgs branch does jump
leading to the noncompact moduli spaces
discussed above in theories with  $N_f \ge 2$.
However, it was shown in \apsw\ that there are no jumps
in the Higgs branch at the superconformal points
where {\it non-mutually  local} singularities collide.
This distinction was called
the ``Higgs branch criterion'' in  \apsw\
and was a key technical tool used to find
the nontrivial superconformal theories in the $SU(2)$ theories with matter.
According to the Higgs branch criterion, the Higgs branch at the $(k,1)$
points is the Higgs branch of
the theory with $k$ hypermultiplets: there is no extra source of
noncompactness when the masses take their critical values.
This shows in particular that, in the $N_f=1$ theory that we are
studying, the moduli space
of vacua will remain compact for $m=m_*$.

{}From a strictly mathematical point of view, we should consider
Theorem 4.3.1 as a
conjecture, but we should also emphasize that the above
result is rigorous at the physical level. As we will see, this
theorem has far-reaching implications for the topology of
four-manifolds: looking at \expans, it is clear that the
analyticity of $F(z)$ will give a strong set of constraints on the
structure of $\CS\CW^{\pm}$ if the overall factor has a negative
exponent. As this exponent depends on the numerical invariants of
the manifold under consideration, we conclude that, for certain
values of these invariants, the basic classes of the manifold and
their SW invariants will have to satisfy certain nontrivial
relations.

This result is a generalization of the constraint $x^2 = 2\chi +
3\sigma$ on basic classes for manifolds of simple type, and shows
that the superconformal points of the ${\cal N}=2$ gauge theories
give predictions on four-manifold topology with a very different
flavor from previous results on the ``geography problem''
(described below). The new constraints reveal a relation between
two {\it a priori} unrelated quantities: the classical topological
invariant $c_1^2 -
\chi_h$, and the diffeomorphism invariants obtained from gauge
theory.

In section 6 we develop the consequences of this physical
theorem. In order to appreciate these we first review, in the
next section, the geography problem, and in particular
what is known about possible values of $c_1^2$ and $\chi_h$.
Since our theorem is not mathematically rigorous we check it in a large number
of examples in section 7.

\newsec{A brief review of the geography of four-manifolds}

The geography problem is one of the most active areas of research
in four-manifold topology. Here we will give a brief review of the results
on geography,
focusing on the most relevant aspects for our work.
An updated review with a list of references can be found in \gompfbook.

\subsec{The geography of four-manifolds}
The numerical invariants of a manifold are topological invariants
given by integers. The simplest examples are the Euler
characteristic $\chi$ and the signature $\sigma$. The geography
problem for four manifolds can be formulated as follows: which
pairs of integers can be realized as $(\chi, \sigma)$ of a smooth
four-manifold $X$, and what is the influence of these values on the
geometry of the four-manifold? To investigate the geography
problem, we can also use the quantities $c_1^2$ and $\chi_h$
introduced in \vari, and this is what we will do to facilitate
comparison with mathematical results. Notice that $\chi_h$ (defined
in \thec\ above) is not necessarily an integer for a generic
four-manifold. The condition that $\chi_h$ be an integer is called
the {\it Noether condition}. In particular, any almost-complex
manifold satisfies it. To analyze the geography problem, one should
take into account that there are simple topological constructions
that can change the value of the numerical invariants. The most
important of these constructions is the blow-up process. The smooth
blow-up of a four-manifold $X$ is simply the connected sum
$\widehat X= X\sharp {\overline {\IC P^2}}$. This construction can
be considered in the complex and the symplectic categories ({\it
i.e.}, one can perform the blow-up in such a way that the resulting
manifolds are complex or symplectic, respectively). Notice that,
under a blowup, the numerical invariants of the four-manifold
change as follows:
\eqn\vari{
c_1^2 (\widehat X) = c_1^2 ( X) -1, \,\,\,\,\,\,\ \chi_h (\widehat
X)= \chi_h (X).} Therefore, we can generate arbitrarily low values
of $c_1^2 $ just by considering a succession of blow-ups. This is
why the study of geography of complex and symplectic four-manifolds
focuses on {\it minimal} four-manifolds. A complex surface is
minimal if it doesn't contain a holomorphically embedded sphere of
square $(-1)$. A minimal symplectic four-manifold can be defined in
a similar way by changing ``holomorphically embedded curve" to
``symplectically embedded sphere" in the above definition. Notice
that a spin complex or symplectic manifold is always minimal: by
the  Wu formula $(w_2(X), \alpha) = \alpha^2$ mod $2$, so if there
is a class with $\alpha^2 = -1$ then $w_2(X) \not=0$.

We have stated the geography problem for general four-manifolds,
but in fact the problem has been studied by restricting to more
specific subsets, where it can be analyzed in more detail, and then
enlarging progressively the set of manifolds under consideration.
The most complete results have been obtained for complex surfaces.
Symplectic manifolds have been also considered in some detail, but
are not classified. We will consider the case of complex surfaces
separately, and then
we will briefly review what is known in the symplectic case, as
well as in more general situations.

\subsec{Geography of complex surfaces}

We will review now some useful facts about the geography of minimal complex
surfaces.
\foot{Since we have no wish to write a textbook on
4-manifolds we use several terms here without
definition. The reader should consult, e.g.,
\gompfbook\GrHa\beau\ or \bpv\ Chapter V, section 5,
pages 146-147
and Chapter VI.}

 Complex surfaces can be classified according to
the Kodaira dimension $\kappa (S)$. This quantity measures
the number of holomorphic $(2n,0)$ forms for large $n$ and can take the values
$-\infty, 0, 1$ and $2$ \bpv\beau. In fact, the Kodaira dimension
already gives important restrictions on the possible values of the
numerical invariants. We have the following possibilities:

\ifig\geoone{The geography of four-manifolds. The various lines are
explained in the text. No known irreducible manifolds lie
below the $\chi_h$ axis. The non-existence of such manifolds is the
``3/2 conjecture.'' No known spin manifolds lie below the ``$11/8$
line.'' The non-existence of such manifolds is the ``$11/8$
conjecture". } {\epsfxsize3.0in\epsfbox{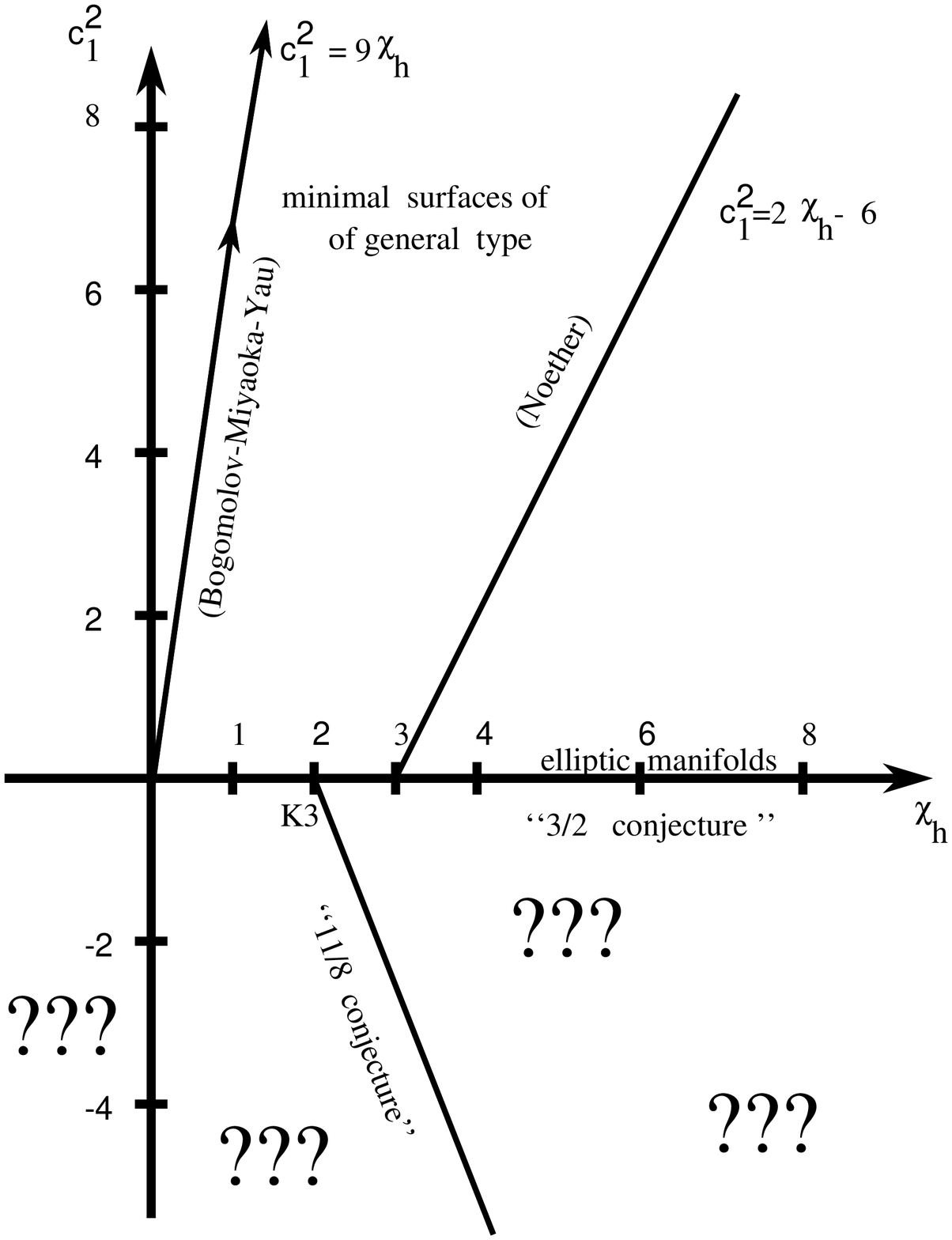}}

$\bullet$ The minimal surfaces with $\kappa (S)=-\infty$ are $\IC
P^2$, geometrically ruled ({\it i.e.} a sphere bundle over a
Riemann surface) or of Kodaira class VII. All these surfaces have
$b_2^+=0$ (for class VII) or $1$. For a geometrically ruled
surface, one has $c_1^2 = 8(1-g)$ and $\chi_h=1-g$. We won't
consider surfaces with $\kappa (S)=-\infty$ in the following, since
they have $b_2^+ \le 1$.

$\bullet$ The minimal surfaces with $\kappa (S)=0$ have $c_1^2=0$
and $\chi_h\ge 0$. In fact, there are only five types of surfaces
in this class: Enriques surfaces, bielliptic surfaces, Kodaira
surfaces (primary and secondary), abelian surfaces (tori), and $K3$
surfaces. Enriques, bielliptic and secondary Kodaira surfaces have
$b_2^+\le 1$. Abelian varieties and primary Kodaira surfaces have
$\chi_h=0$, and $K3$ surfaces have $\chi_h=2$. Again, we will be
only interested in the surfaces with $b_2^+>1$, which in this case
are tori, $K3$ and primary Kodaira surfaces. All of them are in
fact elliptic fibrations, which we consider now.

$\bullet$ An elliptic fibration is a complex surface $S$ together
with a holomorphic fibration $\pi: S \rightarrow \Sigma_g$ over a Riemann
surface of genus $g$, where the generic fibres are elliptic curves.
All the minimal surfaces with Kodaira dimension $\kappa (S)=1$ are
elliptic fibrations, but the converse is not true, as the examples
above show. Any minimal elliptic surface has $c_1^2 =0$ and $\chi_h \ge
0$. Actually, all the nonnegative values of $\chi_h$ are realized.
Therefore, the minimal elliptic fibrations fill the line $c_1^2=0$ in the
$(\chi_h, c_1^2)$ plane (see \geoone).

$\bullet$ The surfaces with Kodaira dimension $\kappa (S)=2$ are
called surfaces of general type. For minimal surfaces of general
type, one has $c_1^2 >0$, $\chi_h >0$. The geography problem for
minimal surfaces of general type has been investigated extensively,
and the following important bounds have been obtained:
 \eqn\minimal{
2 \chi_h -6 \le c_1^2 \le 9 \chi_h.} The lower bound is given by
the {\it Noether line} $c_1^2=2 \chi_h -6$, and the corresponding
inequality is called the {\it Noether inequality}, while the upper
bound corresponds to the {\it Bogomolov-Miyaoka-Yau (BMY)} line. It
has also been proved that all the positive integers $(c_1^2,
\chi_h)$ in the region $2 \chi_h -6 \le c_1^2 \le 8 \chi_h$ are, in
fact, actually realized by minimal surfaces of general type (see
\gompfbook\ for a detailed discussion). The line $c_1^2 = 8 \chi_h$
is called the $0$-signature line. It is not yet
known if all the points in the ``arctic region"
$8 \chi_h \le c_1^2 \le 9 \chi_h$ are inhabited by simply-connected minimal
surfaces of general type.

This is all we will need about complex surfaces. We will consider
now the geography problem for more general manifolds.

\subsec{Geography of symplectic manifolds and beyond}

Symplectic manifolds have played an important role in recent
developments on four-manifold topology. For some time it was
thought that, although the class of symplectic manifolds was
strictly larger than the class of K\"ahler manifolds, the
non-K\"ahler symplectic manifolds were rather special,
comprising perhaps only a few examples in the simply connected case. The new
constructions of symplectic manifolds by Gompf  \gompfnc\ shattered
this view, and now many examples
of non-K\"ahler symplectic manifolds are known. Therefore, one can wonder
about the geography problem for minimal, symplectic manifolds.
Interestingly enough, some of the results have been obtained using
the results by Taubes about the SW invariants of symplectic
manifolds \taubes\ko. This is a simple consequence of Taubes'
results that, for minimal, simply-connected, symplectic
four-manifolds one has $c_1^2 \ge 0$. Moreover, it can be shown
that there are manifolds in this category that violate the Noether
inequality. Indeed, all the values $(c_1^2, \chi_h)$ satisfying $0\le c_1^2
\le 2 \chi_h-6$ are realized by minimal, simply
connected, symplectic manifolds (for some examples of these
manifolds, see \gompfnc).

A natural extension of symplectic manifolds is that of {\it
irreducible} manifolds. A smooth four-manifold is called
irreducible if, for every smooth connected sum decomposition $X=
X_1 \sharp X_2$, either $X_1$ or $X_2$ is homeomorphic to ${\bf
S}^4$. In other words, an irreducible four-manifold is not the
connected sum of non-trivial four-manifolds. It also follows from
Taubes' theorems on SW invariants for symplectic manifolds that a
simply connected, minimal, symplectic four-manifold with $b_2^+>1$
is irreducible.  It was thought for some time that symplectic
manifolds are the building blocks of irreducible, simply connected
manifolds, but once again this view was shattered by the
discovery of many examples of irreducible,
simply-connected manifolds that do not admit a symplectic structure
\szabo\fsknot. Again, SW invariants played a very important role in
these constructions.

Many questions regarding the geography problem
for irreducible four-manifolds have not been solved yet. It has
been conjectured, for example, that the inequality $c_1^2\ge 0$
holds for irreducible manifolds (this conjecture is called the
$3/2$ conjecture). In addition, irreducible manifolds do not
necessarily satisfy the Noether condition. When this condition
does not hold, all the SW invariants are zero and very little is known.

Another conjecture in geography deals with simply connected, spin
manifolds. The intersection form of this class of manifolds is equivalent to
\eqn\qint{
Q= 2k E_8 \oplus \ell II^{1,1},}
where $E_8$ denotes the Cartan matrix of
this Lie algebra, and $II^{1,1}$ is the even unimodular rank two lattice.
Notice that $k$ can be positive or negative, depending
on the sign of the signature $\sigma =16k$. The $11/8$ conjectures
states that  $l\ge 3|k|$. For manifolds with negative signature,
this corresponds to the inequality
\eqn\elei{
c_1^2 \ge {8 \over 3} (2- \chi_h).} This is the line denoted by
``$11/8$" in \geoone. Evidently,    the $3/2$ conjecture implies
the $11/8$ conjecture.
\foot{Actually, there is a small region below the $11/8$ line
but above the $\chi_h$ axis. It is easy to check that all the
integral points in this region are inconsistent with the spin
condition.}

\newsec{New results on the geography of four-manifolds}

In this section, we will use our main result in section 4 (namely,
that the correlation functions are finite) to extract some
interesting information about the geography of four manifolds. We
will find a set of constraints that relate the structure of the SW
invariants and the basic classes to the value of $c_1^2- \chi_h$.
First we will find a sufficient condition for the correlation
functions to be finite and we will introduce the important concept
of {\it manifolds of superconformal simple type}. Then we will
consider more general possibilities for analyticity of $F(z)$ in
some detail. Finally, we study manifolds of simple type with only
one basic class, and we prove (using our main result) that their
numerical invariants must satisfy the inequality $c_1^2 \ge
\chi_h-3$.

\subsec{A sufficient condition for regularity:
manifolds of superconformal simple type}

 In this subsection , we  give a simple
sufficient condition for $F(z)$ to be analytic. To do this, it is
convenient to work with Laurent series in
{\it integral} powers of $z$.

We consider first
\eqn\swone{
\eqalign{
\CS\CW_0 (z) &= z^{ c_1^2 - \chi_h \over 4}
 \biggl(  \CS\CW^+ (z^{1/4}) +
 {\rm e} ^{{\pi i \over 2} (c_1^2-\chi_h)} \CS\CW^{-} (z^{1/4}) \biggr),\cr
\CS\CW_1 (z) & = z^{{ c_1^2 - \chi_h +2 \over 4} } \biggl(  \CS\CW^+ (z^{1/4})
-
{\rm e} ^{{\pi i \over 2} (c_1^2-\chi_h)} \CS\CW^{-} (z^{1/4}) \biggr).\cr}}
Notice that these are precisely the combinations that appear in
\expans. It is easy to prove that these functions have a power series
expansion in integral powers of $z$. There are two cases to
consider, depending on the parity of $\chi_h + \sigma$. If  $\chi_h
+ \sigma$ is even, then $\CS\CW^{\pm}$ have a series expansion in powers
of $z^{1/2}$, as we saw in section 4. But in this case
\eqn\sim{
c_1^2- \chi_h = 7\chi_h + \sigma} is also even. Therefore, $z^{
c_1^2 - \chi_h \over 4} \CS\CW^+ (z^{1/4})$ only contains even
powers of $z^{1/4}$. Under $z^{1/2} \rightarrow -z^{1/2}$, this
function changes as
\eqn\change{
z^{ c_1^2 - \chi_h \over 4} \CS\CW^+ (z^{1/4}) \rightarrow  {\rm e}
^{{\pi i \over 2} (c_1^2-\chi_h)}z^{ c_1^2 - \chi_h \over 4}
\CS\CW^{-} (z^{1/4}).} Therefore, the sum of these two terms only
contains integral powers of $z$, and their difference only contains
half-integral powers of $z$. As $\CS\CW_1$ has an extra power of
$z^{1/2}$, we have proved our claim. The analysis when $\chi_h +
\sigma$ is odd is completely analogous.

In an analogous way we can define
\eqn\nwseries{
\eqalign{
\CA^{(0)} & \equiv \half z^{-(c_1^2-\chi_h)/4} \Biggl( \CA^+
+ {\rm e} ^{- {\pi i \over 2} (c_1^2-\chi_h)} \CA^- \Biggr), \cr
\CA^{(1)} & \equiv  \half z^{-(c_1^2-\chi_h +2)/4  } \Biggl( \CA^+
- {\rm e} ^{- {\pi i \over 2} (c_1^2-\chi_h)} \CA^- \Biggr), \cr
\CA^\pm & \equiv (c^{-1}\kappa_{\pm})^{\chi_h} \biggl[  \bigl(
{du \over da} \bigr) ^2_{\pm}  \biggr] ^{ {\chi_h + \sigma \over 2}
} {\rm e}^{  2 p (u_{\pm}-u_*) + S^2 (T_{\pm}-u_*/3) }. \cr}
}
In terms of these functions we may write
\eqn\anf{
F(z) = \CA^{(0)}(z) \CS\CW_{0 }(z) +
\CA^{(1)}(z) \CS\CW_{1 }(z).
}
The advantage of this representation is that, by
an argument analogous to that for $\CS\CW_{0 },\CS\CW_{1 }$
we see  that $\CA^{(0)},\CA^{(1)}$ have power
series in integral {\it nonnegative}  powers of $z$:
\eqn\nwserii{
\eqalign{
\CA^{(0)} & = \sum_{k=0}^\infty A_k^{(0)} z^k, \cr
\CA^{(1)} & = \sum_{k=0}^\infty A_k^{(1)} z^k, \cr}
}
where the coefficients are polynomials in
$p$, $S^2$, $\chi_h $ and $c_1^2$. For instance, working
at next-to-leading order in the expansions in \scddone, we find
\eqn\firstco{
A_0^{(0)}=1, \,\,\,\,\,\,\,\,\,\ A_0^{(1)}= \bigl({ 4 \over 3}
\bigr)^{4/3}\Bigl(  { c_1^2 + 5 \chi_h \over 2} \Bigr) -{S^2 \over
24 },} corresponding to the first terms written in \expans. Of
course, the next terms become increasingly complicated.


It would be
extremely useful to have a precise criterion  for the regularity of
$F(z)$ at $z=0$. It is clear from \anf\ that, if $\CS\CW_0 (z)$,
$\CS\CW_1 (z)$ are regular at $z=0$, then $F(z)$ will be regular as
well.  We will now state a theorem that gives a simple condition
for $F(z)$ to be regular. In this theorem, the basic classes are
regarded as functionals acting on $H^2 (X;\IZ)$ through the
intersection form: $\lambda (S) = (\lambda, S)$. More generally,
the functionals $\lambda^n$ act on ${\rm Sym}^n (H^2 (X; \IZ))$ as
follows:
\eqn\dual{
\lambda^n (S_1 \cdots S_n) = (\lambda, S_1)\cdot \dots \cdot (\lambda, S_n).}

\bigskip
\noindent
{\bf Theorem 6.1.1}.

a.)  If $\chi_h - c_1^2 -4 < 0$ ( i.e.,  $c_1^2\ge \chi_h-3$) then
$F(z)$ is  regular at $z=0$.

b.) If $\chi_h - c_1^2 -4 \ge 0$ and the following relations are
satisfied
\eqn\sumrules{
\sum_{\lambda} SW (\lambda) {\rm e}^{ 2\pi i (\lambda_0^2 + \lambda \cdot
\lambda_0)}
 \lambda^{k} =0, \,\,\,\,\,\,\,\ k=0, \dots,  \chi_h - c_1^2 -4, }
then the function $F(z)$ is regular at $z=0$.

\bigskip

Notice that, if $\chi_h + \sigma$ is even (odd), the expressions of the
form \sumrules\
with $k$ odd (even) are automatically zero. Therefore,
there are actually $(\chi_h - c_1^2 -4)/2 +1$ nontrivial
conditions if $\chi_h + \sigma$ is even, and $(\chi_h - c_1^2 -3)/2$
nontrivial conditions if $\chi_h + \sigma$ is odd.

\bigskip
\noindent
{\it Proof}: If condition $(a)$ holds then the leading
power of $z$ in
$\CS\CW_0 (z)$ and $\CS\CW_1 (z)$
is greater than or equal to $z^{-3/4}$. Since these
series have no monodromy, they must be regular.
Therefore we may assume condition $(b)$ holds.

To prove the theorem, we will show that the condition \sumrules\ is
equivalent to the regularity of $\CS\CW_0 (z)$ and $\CS\CW_1 (z)$
at $z=0$. To do this, we have to consider the different values of
$c_1^2-\chi_h$. If $c_1^2 - \chi_h$ is even, then $\chi_h + \sigma$ is
even, and  $\CS\CW^{+}
(z^{1/4})$ has the expansion
\eqn\expfour{
\CS\CW^+ (z^{1/4})= \sum_{n=0}^{\infty} a_{2n} z^{n/2}. }
If $c_1^2- \chi_h$ is odd, $\chi_h + \sigma$ is odd as well and
one has
\eqn\anexp{
\CS\CW^+ (z^{1/4})= \sum_{n=0}^{\infty} a_{2n+1} z^{(2n+1)/4}.}
The structure of the series $\CS\CW_0 (z)$ and $\CS\CW_1 (z)$
depends on the value of $c_1^2 - \chi_h$ mod $4$. We then have four different
cases:

1) Suppose that $c_1^2- \chi_h=4r$.
By condition $(b)$, $r$ is a negative integer. Using the definitions
\swone, one finds
\eqn\czero{
\CS\CW_0 (z) = 2 \sum_{p=0}^{\infty} a_{4p} z^{p + r},
 \,\,\,\,\,\,\,\  \CS\CW_1 (z) = 2 \sum_{p=0}^{\infty} a_{4p+2} z^{p + r +1}.}
It follows that $\CS\CW_0$ and $\CS\CW_1$ are regular at $z=0$ if
and only if
$$
a_{2n}=0 \qquad {\rm  for } \qquad 0 \leq 2n \le \chi_h - c_1^2 -4 = 4 (\vert r
\vert -1).
$$

2) If $c_1^2- \chi_h=4r + 2$, $r<-1$ by condition $(b)$. In this
case, one finds:
\eqn\ctwo{
\CS\CW_0 (z) = 2 \sum_{p=0}^{\infty} a_{4p+2} z^{p + r+1},
 \,\,\,\,\,\,\,\  \CS\CW_1 (z) = 2 \sum_{p=0}^{\infty} a_{4p} z^{p + r +1}.}
Again, we find that $\CS\CW_0$ and $\CS\CW_1$ are regular at $z=0$ if
and only if $a_{2n}=0$ for $2n \le \chi_h - c_1^2 -4
= 4 \vert r \vert - 6$.

3) If $c_1^2- \chi_h=4r + 1$, $r<-1$ by condition $(b)$. The series
$\CS\CW_0 (z)$ and $\CS\CW_1 (z)$ have the form
\eqn\cone{
\CS\CW_0 (z) = 2 \sum_{p=0}^{\infty} a_{4p+3} z^{p + r+1},
 \,\,\,\,\,\,\,\  \CS\CW_1 (z) = 2 \sum_{p=0}^{\infty} a_{4p+1} z^{p + r +1}.}
These functions are regular if and only if $a_{2n+1} =0$ for $2n+1 \le
\chi_h - c_1^2 -4= 4 \vert r \vert - 5 $.

4) Finally, if $c_1^2- \chi_h=4r + 3$, condition $(b)$ imposes
again $r<-1$. The series $\CS\CW_0 (z)$ and $\CS\CW_1 (z)$ are now
\eqn\cone{
\CS\CW_0 (z) = 2 \sum_{p=0}^{\infty} a_{4p+1} z^{p + r+1},
 \,\,\,\,\,\,\,\  \CS\CW_1 (z) = 2 \sum_{p=0}^{\infty} a_{4p+3} z^{p + r +2}.}
The necessary and sufficient condition for regularity of these
series at $z=0$ is again that $a_{2n+1} =0$ for $2n+1 \le
\chi_h - c_1^2 -4= 4 \vert r \vert - 7 $.

In all the cases, we find that $\CS\CW_0$ and $\CS\CW_1$ are
regular at $z=0$ if and only if $a_k=0$ for  $k\le \chi_h - c_1^2
-4$, and $k$ even (odd) for $\chi_h + \sigma$ even (odd). In order
to relate the coefficients $a_k$ in the expansions \expfour\anexp\
to the SW invariants, we have to expand $(du/da)^2_{+}$ in powers
of $z^{1/2}$:
\eqn\expdu{
\bigl( {du \over da} \bigr)^2_{+}=z^{1/2} (1 + b_1 z^{1/2} + \dots),  }
where $b_1 = (4/3)^{4/3}$. One finds that
\eqn\coeffa{
a_k = {(-i)^k \over k!} \sum_{\lambda} SW (\lambda) {\rm e}^{ 2\pi
i (\lambda_0^2 + \lambda \cdot \lambda_0)}(\lambda, S)^{k} +
p(a_{k-2},\dots), } where $p(a_{k-2}, \dots)$ is a linear function
of the previous $a_{i}$, $i<k$, $i\equiv k$ mod $2$, whose
coefficients depend on the coefficients in the expansion \expdu.
For example,
\eqn\coeffsfe{
\eqalign{
a_0 &= \sum_{\lambda} SW (\lambda){\rm e}^{ 2\pi i (\lambda_0^2 +
\lambda \cdot \lambda_0)}, \,\,\,\,\,\,\,\ a_1   = -i\sum_{\lambda}
SW (\lambda) {\rm e}^{ 2\pi i (\lambda_0^2 + \lambda \cdot
\lambda_0)}(\lambda, S),
\cr
a_2 &= -{1 \over 2}\sum_{\lambda} SW (\lambda) {\rm e}^{ 2\pi i
(\lambda_0^2 + \lambda \cdot \lambda_0)}(\lambda,
S)^{2},\,\,\,\,\,\,\,\ a_3  = {i \over 3!} \sum_{\lambda} SW
(\lambda) {\rm e}^{ 2\pi i (\lambda_0^2 + \lambda \cdot
\lambda_0)}(\lambda, S)^3 + {b_1 \over 2} a_1,\cr a_4&={1 \over 4!}
\sum_{\lambda} SW (\lambda) {\rm e}^{ 2\pi i (\lambda_0^2 + \lambda
\cdot \lambda_0)}(\lambda, S)^{4} + b_1 a_2,\cr }} and so on.
Clearly, $a_{k}=0$ for $k \le \chi_h - c_1^2 -4$ if and only if
\eqn\lastcon{
\sum_{\lambda} SW (\lambda) {\rm e}^{ 2\pi i (\lambda_0^2 +
\lambda \cdot \lambda_0)}(\lambda, S)^{k}=0,
 \,\,\,\,\,\,\ k=0, \dots, \chi_h - c_1^2 -4,}
and this must be true for every $S \in H^2 (X; \IZ)$. As the
intersection form is nondegenerate, we obtain the condition
\sumrules\ stated in the theorem. $\spadesuit$

Therefore, we see that, if a manifold of simple type satisfies the
conditions \sumrules, regularity of $F(z)$ will be automatically
guaranteed, without any further knowledge of the expansions of the
physical quantities around the superconformal point. We stress that
\sumrules\ is only a sufficient condition for regularity, and we
will see in the next section that there are other possibilities for
$F(z)$ to be regular. These possibilities
are much   more complicated and  involve
next-to-leading terms in the  expansion in $z$.
The
simplicity of the condition \sumrules\ suggests the following

\bigskip
\noindent
{\bf Definition 6.1.2}. Let $X$ be a compact, oriented manifold with
$b_2^+>1$ and of simple type. We say that $X$ is of {\it
superconformal simple type} if it satisfies any of the conditions
stated in the previous theorem, {\it i.e.} if $c_1^2\ge \chi_h-3$
or  $\chi_h - c_1^2 -4 \ge 0$ and the relations \sumrules\ hold.
\bigskip

\rmk{}

\item{1.} Notice that any manifold with trivial SW invariants is of
superconformal simple type.

\item{2.} The concept of a manifold of superconformal simple type has some
resemblances with the concept of manifold of simple type. From the
results of \monopole\mw, it has become clear that a manifold is of
simple type if, in the expansion of the different quantities around
the monopole singularity, only the leading term is relevant.
Similarly, a manifold is of superconformal simple type if, in the
analysis of regularity around the superconformal point, only the
leading term is relevant. This means in particular that the
constraints on the geometry of the manifold that follow from the
regularity of $F(z)$ are dictated only by the universal behavior of
the critical theory ({\it i.e.} by the anomalous dimensions of the
operators). Notice that the quantum field theory analysis does not
imply that a manifold with $b_2^+>1$ is of simple type, and in the
same way our result that the generating function is regular at the
superconformal point does not imply that a manifold is of
superconformal simple type.
\foot{However, from the QFT viewpoint the conjecture
that all manifolds of $b_2^+>1$ are of simple type is
rather natural.}

\item{3.} From the mathematical point of view, there are also some
similarities between the two concepts. We will show in section 7
that all complex surfaces are of superconformal simple type, and
that this property is preserved under blowup and other standard
constructions. Indeed, we haven't found any example of a manifold
of $b_2^+>1$ and of simple type which is not  of superconformal
simple type.

\item{4.} An alternative characterization of manifolds of superconformal
simple type is the following. For any manifold $X$ with $b_2^+>1$
and of simple type, one can define the following SW series with
magnetic flux $w_2 (E)=w_2 (X)$:
\eqn\swserwz{
SW_X^{w_2 (X)}  = \sum_{x} (-1)^{c_0^2 + c_0\cdot x \over 2} SW (x)
{\rm e}^{ x}, } where the notations are as in section 4.1. As in
\sumrules, the exponential ${\rm e}^x$ is understood here as a
multilinear map on ${\rm Sym}^{*} (H^2 (X, \IZ))$. Notice that
\swserwz\ is {\it not} the usual SW series considered in the
mathematical literature (see, for example, \fsknot), which is
defined with zero magnetic flux. According to the result of Witten
\monopole, the  series \swserwz\ is related to the Donaldson series for $w_2
(E)=w_2 (X)$ as follows:
\eqn\dseries{
{\CD}_X^{w_2(X)} = 2^{2 + c_1^2 -\chi_h} {\rm e}^{Q/2} SW_X^{w_2
(X)},} where $Q$ is the intersection form. Consider now the
following holomorphic function of $z$, obtained from the SW series
after replacing  $x \rightarrow z x$  :
\eqn\swseries{
SW_X^{w_2 (X)} (z) = \sum_{x} (-1)^{c_0^2 + c_0\cdot x \over 2} SW
(x) {\rm e}^{z  x}.} If we expand this holomorphic function around
$z=0$, we find
\eqn\expand{
SW_X^{w_2 (X)} (z) = \sum_{n=0}^{\infty}
\biggl( \sum_{x}  (-1)^{c_0^2 + c_0\cdot x \over 2}
SW (x) x^n \biggr) {z^n \over n!}.}
According to our definition, $X$ is of superconformal simple type
if $c_1^2 \ge \chi_h -3$ or $\chi_h -c_1^2 -4 \ge 0$ and the
first $\chi_h -c_1^2 -4$ coefficients of \expand\ are zero.
Therefore,

\bigskip
\noindent
{\bf Proposition 6.1.3}. $X$ is of superconformal simple type if and only if
$SW_X^{w_2 (X)} (z)$ has a zero at $z=0$ of order $\ge
\chi_h - c_1^2 -3$.

\bigskip

Notice that, depending on the parity of
$\chi_h + \sigma$, we will have even or odd powers of $z$ in
\expand. If a manifold is of superconformal simple type and
$\chi_h -c_1^2 -4 \ge 0$, the order of the zero
in the series \swseries\ is in fact greater or equal than
$\chi_h -c_1^2 -2$.
Proposition 6.1.3 will be very useful in section seven: given a manifold of
$b_2^+>1$ and of simple
type, to check that it is of superconformal simple type
we only have to compute the SW series \swseries\ and
examine the order of vanishing at $z=0$. Notice that,
in \swseries, $z$ is a formal
variable, different from the physical expansion parameter
considered before, although it plays a similar role from the point
of view of the analyticity. As we will see in section 7, the sign
in \swseries\ depending on the second Stiefel-Whitney class is
crucial. It is interesting that the analysis of the Donaldson
series as a series in the holomorphic variable $z$ is one of the
key steps in the proof of the structure theorem of Kronheimer and
Mrowka \kmone.

\subsec{Examples of more general conditions}

As we have remarked, the conditions \sumrules\ are not the most
general conditions to achieve regularity of $F(z)$. In this
section, we will briefly examine the general condition for
regularity of $F(z)$ that can be derived from the expression \anf.
We just expand this expression in powers of $z$ in the usual way
and write the conditions derived from it. The exact form of the
conditions depends on the residue class of $c_1^2 - \chi_h$ mod 4.
We will only consider in some detail the case of $c_1^2- \chi_h
=4r$. The other cases are similar and can be easily worked out.

If $c_1^2- \chi_h =4r$, the series $\CS\CW_0$, $\CS\CW_1$ have the
structure given in \czero. Using \anf,
we find that $F(z)$ has the following expansion:
\eqn\expmodfour{
F(z) = 2 a_0 \sum_{k=0}^{\infty} A_k^{(0)} z^{r + k} +
2\sum_{k=0}^{\infty} \biggl( \sum_{p=0}^{k} A_{k-p}^{(0)} a_{4p +
4} + A_{k-p}^{(1)} a_{4p +2}\biggr) z^{r + k +1}.} Suppose that
$r<0$. Then, as $A_0^{(0)} =1$, if $F(z)$ is regular we necessarily
have
\eqn\firstcon{
a_0 = \sum_{\lambda} SW (\lambda) {\rm e}^{ 2\pi i (\lambda_0^2 +
\lambda \cdot \lambda_0)}=0.} If $r<-1$, the next conditions are
more involved. It follows from \expmodfour\ that we need
\eqn\gencon{
 \sum_{p=0}^{k} A_{k-p}^{(0)} a_{4p + 4} + A_{k-p}^{(1)} a_{4p +2} =0,
  \,\,\,\,\,\ k=0, 1, \dots, |r|-2.}
If the manifold is of superconformal simple type, the coefficients
$a_{4p}$ and $a_{4p+2}$ will vanish separately for $p \leq \vert k
\vert -2 $, but in principle  we could have cancellations between
the coefficients with indices $0$ mod $4$ or $2$ mod $4$. It is in
fact instructive to write the first condition in \gencon,
corresponding to $k=0$, $r<-1$. We have:
\eqn\genfirst{
a_4 + A_0^{(1)} a_2=0, } where $a_4$, $a_2$ are given explicitly in
\coeffsfe. Notice that in analyzing these equations, we have to
consider the terms with the same powers of $S$. Therefore, the
equation \genfirst\ contains in fact the two equations,
\eqn\reallytwo{
\eqalign{
\biggl( {c_1^2 + 5 \chi_h \over 2} + 1 \biggr)
 \sum_{\lambda} SW (\lambda)
 {\rm e}^{ 2\pi i (\lambda_0^2 + \lambda \cdot \lambda_0)}
 (\lambda, S)^{2} =0,\cr
2 \sum_{\lambda} SW (\lambda) {\rm e}^{ 2\pi i (\lambda_0^2 +
\lambda \cdot \lambda_0)}(\lambda, S)^{4} +S^2 \sum_{\lambda} SW
(\lambda) {\rm e}^{ 2\pi i (\lambda_0^2 + \lambda \cdot
\lambda_0)}(\lambda, S)^{2}=0.\cr }} The first equation has two
possible solutions: either $c_1^2 + 5 \chi_h+2=0$, or the second factor
(which is one of the sum rules we have found in \sumrules)
vanishes. If the latter condition is true, then the second
equation tells us that we also have the sum rule of \sumrules\
for $k=4$. Notice that, in the first equation, if the second
factor is different from zero but the first vanishes, we still have
to satisfy the second equation, that involves a constraint in
the intersection form of the manifold. The general picture follows
this pattern: for each of the equations in \gencon, {\it either}
we have a set of simple sum rules like \sumrules, {\it or} we
have conditions involving $c_1^2$, $\chi_h$ and the intersection
form of the manifold, and probing the higher order terms in the
expansion of the physical quantities. The same situation holds
for the other values of $c_1^2 -\chi_h$ mod $4$.

Fortunately, in the concrete manifolds that we have analysed, the
more general (and complicated) possibilities for the regularity of
$F(z)$ do not play any role: as we spell out in the next
section, all known 4-manifolds   are of superconformal
simple type. We should stress, however, that regularity of $F(z)$
is {\it not} equivalent to the superconformal simple type
condition: as we have seen, more general possibilities are allowed.
The analysis of these possibilities depends very much on the
manifold under consideration, but in some simple cases it can be
done in detail. We will consider such an example in the next
section, which also illustrates the general conditions we have been
discussing.

\ifig\geotwo{The line $c_1^2 = \chi_h -3$ and the geography of
four-manifolds. The manifolds under this line have to satisfy
sum rules for the SW invariants. }
{\epsfxsize3.0in\epsfbox{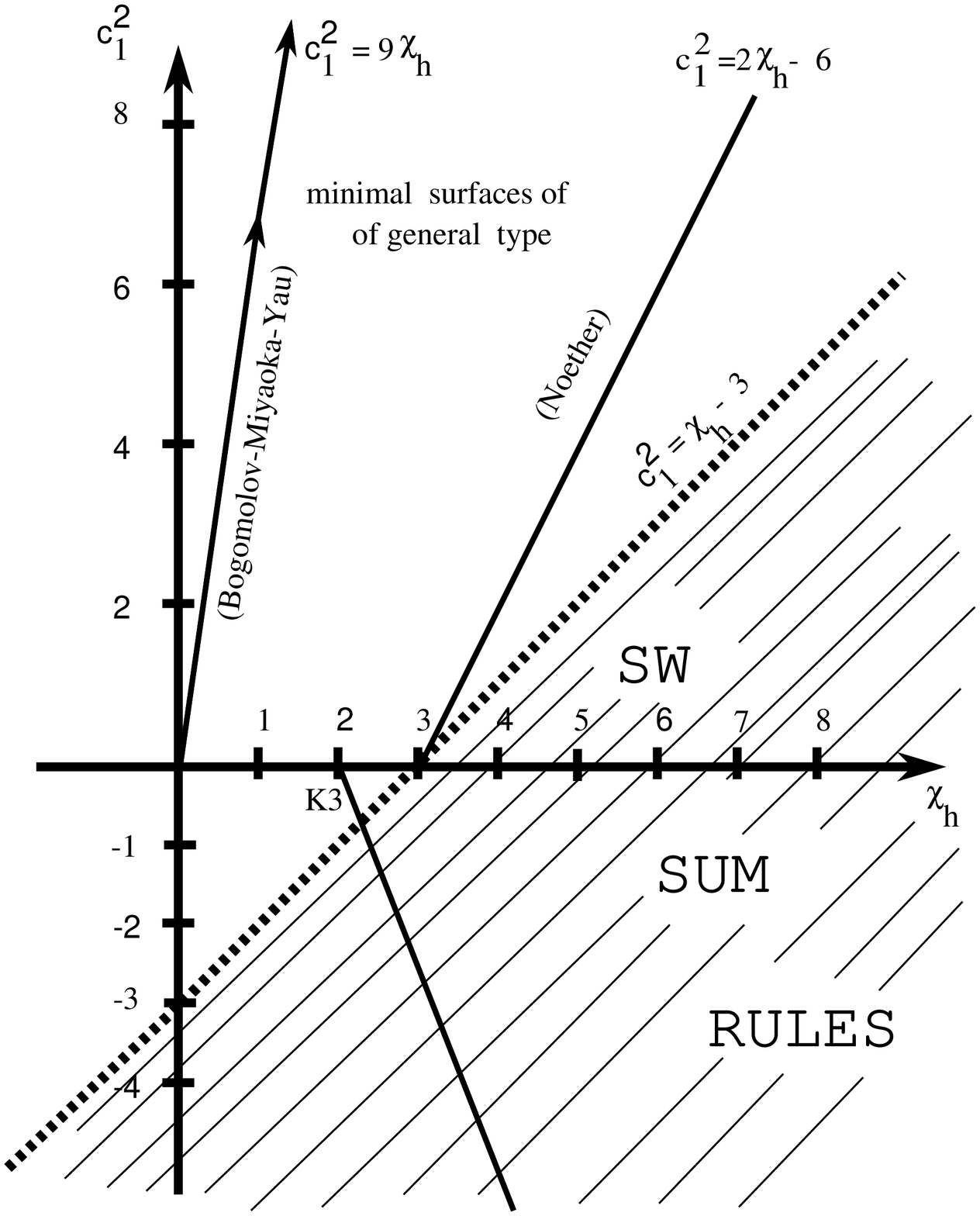}}

At this point, we can discuss the results we have obtained so far
in the context of the geography of four-manifolds. Our analysis
shows that there is a special line in the $(\chi_h, c_1^2)$  plane
that separates two well-distinguished regions (see figure 2). For
$c_1^2 \ge \chi_h-3$, the function $F(z)$ is always regular, as we
have seen in section 6.1. When $c_1^2\le \chi_h-4$, our result
about the regularity of $F(z)$ implies a series of sum rules for
the SW invariants. These rules can take a very simple form, as in
\sumrules, or follow more complicated patterns.  It is interesting
that the region where our constraints are non trivial is precisely
the most intriguing from the point  of view of geography. For
instance, our results put severe constraints on the SW invariants
and basic classes of possible irreducible manifolds with $c_1^2<0$,
or possible spin manifolds that violate the $11/8$ conjecture.

As far as we know, the significance of this particular combination
of numerical invariants, $c_1^2 -\chi_h =(7\chi+ 11 \sigma)/4$, has
not been discussed in the mathematical literature. However, this
quantity does show up as an ``experimental" bound to construct
manifolds with only one basic class. We will explain why
this is so in the next section. Also, this invariant appears in
the famous expression for the Donaldson series due to Witten
\wittk\monopole\ (see \dseries), and very much for the same
reasons, since it enters through the $\chi$, $\sigma$-dependent
factors in the measure of the twisted theory.

\subsec{Manifolds with one basic class}

The purpose of this section is to extract a very concrete
prediction from the regularity of $F(z)$. It follows from \rel\
that, if $x$ is a basic class, then so is $-x$. For this reason, if
$\CB_X$ denotes the set of basic classes of $X$, we say that $X$
has $B$ basic classes if the set $\CB_X/\{ {\pm1}\}$ consists of
$B$ elements. There are many examples of manifolds with only one
basic class, for example the minimal surfaces of general type.
There also examples of noncomplex and nonsymplectic manifolds with
only one basic class \fsrat\fsconst\fsonly. The examples
constructed with this property satisfy $c_1^2 \ge \chi_h-3$, but as
far as we know there is not a clear relation between this bound and
the existence of only one basic class in the manifold. Here we will
prove that, as a consequence of the regularity of $F(z)$, one has
the following
\bigskip
\noindent
{\bf Theorem 6.3.1}.  Let X be a smooth, compact, oriented four-manifold
with $b_2^+>1$ and of simple type. If $X$ has one basic class, then
\eqn\theo{
c_1^2 (X) \ge \chi_h (X) -3.}

\bigskip

\noindent
{\it Proof}: To prove this theorem, we will show that, if $X$ has one basic
class and $c_1^2 (X) < \chi_h (X) -3$, then $F(z)$ cannot be
regular, contradicting our physical theorem. We will use in fact
the general conditions for regularity derived from \anf. In this
case, the analysis is relatively easy because the quantities
involved in the general sum rules are simple. Assume then that $X$
has only one basic class. There are two different cases; the case
when $\chi_h + \sigma$ is even, and the case when $\chi_h + \sigma$
is odd. In the first case, $c_1^2 - \chi_h$ can be $0$ or $2$ mod
$4$, and in the second case it can be $1$ or $3$ mod $4$. We will
denote the only basic class of $X$ by $K$, therefore $SW(K) \not=0$
by assumption. We can choose the integral lifting $c_0= K$. Notice
that, if $K=0$, then both $\chi_h$ and $\sigma$ must be even, due
to \rel\ and to the fact that $K^2=2\chi + 3 \sigma$. As in section
6.1, we have to consider four different cases:

1) Suppose that $c_1^2 - \chi_h =4r$, and that $r<0$. In this case,
according to \firstcon, we must have $a_0=0$. But
\eqn\anut{
a_0= 2(-1)^{\sigma} SW(K) } if $K\not=0$, and $a_0=SW(0)$ if $K=0$.
Hence, $a_0 \not=0$ and we get a contradiction.

2) Suppose that $c_1^2 - \chi_h =4r+2 $ and $r<-1$.
The condition analogous to \genfirst \ is now $A_0^{(1)} a_0 +
a_2=0$. As we discussed in the previous section, this equality
gives in fact two independent equations, and one of them involves
the intersection form. Using the explicit expressions
\coeffsfe\firstco\ and taking into account that $SW(K) \not= 0$, we
find   equations analogous to \reallytwo:
\eqn\inter{
c_1^2 + 5 \chi_h =0, \,\,\,\,\,\, S^2 + 3 (K,S)^2=0, } where we
have taken into account that $K = 2 \lambda$. The second equation
in \inter\ is incompatible with unimodularity of the intersection
matrix because $b_2^+>1$, so $b_2 > 1$ so the intersection form is
more than one-dimensional. Therefore, we find again a
contradiction.

3) Suppose now that $c_1^2 - \chi_h =4r+1 $.
If $r<-1$, one finds that regularity of $F(z)$ implies that
$A_0^{(1)}a_1 + a_3=0$. This gives again two equations,
\eqn\intertwo{
(c_1^2 + 5 \chi_h +1)(K,S)=0 , \qquad   S^2 (K,S) +  (K,S)^3=0, } where both
equations  hold for any $S$. Since
$\chi_h + \sigma$ is odd, $K\not=0$, so, by the
nondegeneracy of the intersection form,
$(c_1^2 + 5 \chi_h +1)=0$. Now consider the
second equation. We can put
$S=K$, and take into account that $K^2 = c_1^2$. We then find that
$c_1^2=0$ or $-1$. If $c_1^2=0$, the first equation has no solution
for an integral $\chi_h$. If $c_1^2=-1$, the first equation gives
$\chi_h=0$, but this contradicts our assumption $c_1^2-
\chi_h=4r+1$. Therefore, \intertwo\ has no solution and we find
again a contradiction.

4) The case $c_1^2- \chi_h=4r +3$ is similar to the first one. If
$r<-1$, one finds as a necessary condition for analyticity of
$F(z)$
\eqn\lastreg{
a_1 = -i (-1)^{\sigma} (K,S) SW(K)=0, } for any $S$. Since $K\not
=0$ (because $\chi_h + \sigma$ is odd in this case), and the
intersection form is nondegenerate, we find again a contradiction.
This ends the proof. $\spadesuit$

This theorem proves that the lower bound for $c_1^2$ that has
been found for manifolds with one basic class
is in fact sharp. Therefore, the examples on the line $c_1^2 = \chi_h -3$
saturate the bound (these examples include $E(3)$ and the manifolds
$Y(n)$ constructed in \fsrat, which will be considered in some
detail in section 7.4).

A corollary of this theorem is that, if a manifold of simple type
with $b_2^+>1$ has only one basic class, it is necessarily of
superconformal simple type. In other words, we have seen that the
more general conditions for regularity of $F(z)$ cannot be
achieved. In fact, we suspect that this will also be the case for
other manifolds, although, as we have seen in this simple case, the
analysis of the conditions is rather delicate.

\newsec{All available 4-manifolds of $b_2^+>1$ are of
superconformal simple type}

In section 4, we have argued that, on physical grounds, the
generating function of the theory near the superconformal $(1,1)$
point has to be regular on every compact four-manifold with
$b_2^+>1$ and of simple type. This leads to some nontrivial
constraints relating the value of the numerical invariant $c_1^2
-\chi_h$ to the SW invariants and the structure of the basic
classes. This is of course a strong statement, and the consequences
are rather surprising. In this section, we will put our result to
the test by actually {\it checking} that it is true for a large
family of four-manifolds. We think that the analysis in this
section provides compelling evidence for our result.

The strategy is the following: first, we will consider minimal
complex surfaces. The analysis of this class of manifolds is fairly
systematic due to the Kodaira-Enriques classification.  We will
show that {\it any} minimal compact complex surface with $b_2^+>1$
is in fact of superconformal simple type. This will be done in
sections 7.1 and 7.2. As we will see, due to the results in section
6, the only nontrivial check is for elliptic fibrations. In section
7.3, we show that the blowup of a manifold of superconformal simple
type is also of superconformal simple type. This, together with the
results in sections 7.1 and 7.2, proves that actually any complex
surface with $b_2^+>1$ (minimal or not) is of superconformal simple type.
In sections 7.4, 7.5 and 7.6 we analyze three different topological
procedures to construct four-manifolds: rational blowdowns, fiber sums
along tori, and knot surgery. As we will see, these three procedures
preserve the superconformal simple type condition. This result is
very interesting, because most of the exotic constructions of four-manifolds
(symplectic and non-symplectic) that we are aware of use these constructions,
and take as their building blocks complex surfaces. Therefore,
without further ado, we can state that all the manifolds constructed using
these operations and starting with manifolds of superconformal simple
type will also be of superconformal simple type. The detailed
examination of these constructions will show to the skeptical
reader the ``magic" of the superconformal simple type condition:
in all the cases, one finds a remarkable balance between the
value of $c_1^2 - \chi_h$ and the order of the zero of the SW series
\swseries. In section 7.7, we consider an ``exotic" example: a family
of symplectic manifolds under the Noether line. These manifolds
are not complex, and are
clearly in a region where the sum rules hold (see \geotwo). We
analyze in some detail a family of manifolds filling the wedge $0\le
c_1^2 \le 2 \chi_h -6$. Although these manifolds are of superconformal
simple type due to the general results about fiber sum along tori
and rational blowdown, the construction is a good example of the uses
of these nice topological constructions. Finally, in section 7.8,
motivated by these results, we state the conjecture that any manifold
of $b_2^+>1$ and of simple type is of superconformal simple type.

\subsec{Minimal surfaces of general type}

 We want to consider minimal complex surfaces, and see
 if they satisfy our physical theorem or not.
If we come back for a while to the Kodaira-Enriques classification,
summarized in section 5.2, we see that the only minimal models
which possibly have $b_2^+>1$ are elliptic fibrations (including
the complex surfaces with $\kappa (S)=0$ and $b_2^+>1$ listed in
section 5.2) and minimal surfaces of general type. Both of them are
of simple type.

We first consider minimal surfaces of general type. As we have
remarked in section 5.2, a classical  result in geography states
that $c_1^2>0$ and $2 \chi_h -6 \le c_1^2$. In particular, $c_1^2
\ge \chi_h -3$ (see \geotwo), therefore all the minimal complex
surfaces are of superconformal simple type, according to our
results in section 6. Notice that minimal surfaces of general type
only have one basic class up to sign, the canonical bundle $K$. It
is a very happy fact that the Noether line turns out to be {\it
above} the boundary line for manifolds with only one basic class
that we found in section 6.3!

\subsec{Elliptic fibrations}
 In many aspects, elliptic fibrations are the
canonical examples for our result, because they have $c_1^2=0$ and
arbitrarily large $\chi_h$. If our result is true, the structure of
the SW series for these surfaces is highly constrained, and we will
find that this is indeed the case. We will then explain in some
detail how to characterize these manifolds topologically, and we
will present the results for the structure of their basic classes
and SW invariants. This material is covered in detail in
\fmbook\fm\brussee. We will focus on relatively minimal
elliptic surfaces, {\it i.e.}, elliptic surfaces with no
exceptional spheres of self-intersection $-1$
in the fibers. All relatively minimal
elliptic surfaces are minimal, except for the blowup of
$\IC P^2$ at nine points.

Recall that an elliptic fibration is a complex surface $S$ together
with a holomorphic fibration $\pi: S \rightarrow \Sigma_g$. Given an elliptic
fibration, we can associate to it a line bundle $L$ over the
Riemann surface $\Sigma_g$ with the property that ${\rm deg}(L) =
\chi_h \ge 0$.
\foot{
For details, see \fmbook, section 1.3.5. Technically $L=( R^1\pi_*
\CO_S)^{-1}$. }
 In the fibration we will have a simple fiber $f$ as well as
 $r$ multiple fibers $f_i$, $i=1, \dots, r$ of multiplicity $m_i$,
 and with the following relation in (co)homology: $f= m_i f_i$. The
 canonical bundle of $S$ can be written in terms of the canonical
 bundle of the Riemann surface,
the line bundle $L$, and the holomorphic line bundles associated to
the multiple fibers. The general expression is
\eqn\canon{
K_S = \pi^{*} (K_{\Sigma_g} \otimes L) + {\cal O}_S \bigl( \sum_i
(m_i-1) f_i \bigr), } which has first Chern class
\eqn\firstchern{
 c_1 (K_S) = (\chi_h + 2g-2) f +  \sum_i (m_i-1) f_i , }
and this gives $c_1^2 (S)=0$ (since $f^2=0$). Therefore, $\chi=12
\chi_h$. There are two different cases for the study of elliptic
fibrations: elliptic fibrations with nonzero Euler number
(equivalently, with $\chi_h>0$) and elliptic fibrations with zero
Euler number. Elliptic fibrations with $\chi_h=0$ and $b_2^+>1$ are
automatically of superconformal simple type (these include tori and
primary Kodaira surfaces). Therefore, we can focus on elliptic
fibrations with $\chi_h>0$, {\it i.e.}, with positive Euler number.
In this case, one has $b_1 (S)=2g$ \fmbook. Another useful
numerical invariant is the geometric genus $p_g (S)$, which for
elliptic fibrations with $\chi_h>0$ is given by \eqn\geogen{ p_g
(S) = \chi_h+g-1.} On the other hand, a well-known theorem by
Kodaira \fmbook\bpv\ states that $b_2^+ (S)=2p_g (S)+1 $ if $b_1
(S)$ is even, therefore $b_2^+ (S)>1$ if and only if $p_g \ge1$.

Now, we can analyze the SW invariants and the basic classes for
minimal elliptic surfaces, that have been worked out in \fm\brussee.
The basic classes
have the form
\eqn\basef{
x_{d, a_i} = (\chi_h + 2g-2 - 2d) f + \sum_i (m_i -1-2a_i) f_i,
\,\,\,\,\,\,\  0\le d \le \chi_h + 2g-2, \,\,\,\,\,\ 0 \le a_i \le
m_i-1,} and the corresponding SW invariants are
\eqn\swef{
SW ( x_{d, a_i} ) = (-1)^d { \chi_h + 2g-2 \choose d} .} Notice
that, since $p_g \ge 1$ and $g \ge 0$, we always have $\chi_h +
2g-2 \ge 0$. We want to compute the SW series \swseries. As
$c_1(K_S) \equiv w_2 (X)$ mod $2$, we can choose the integral
lifting $c_0 = c_1 (K_S)$. As we noticed in \factor, another choice
of the lifting will give an overall $\pm1$ in the series, therefore
it won't change the analyticity property that we want to verify.
With this choice of the lifting, the series \swseries\ is just the
usual one and we find the holomorphic function
\eqn\holef{
SW^{w_2(X)}_X (z) = 2^{\chi_h + 2g-2} \bigl( \sinh (z f)) ^{\chi_h
+ 2g-2} \prod_{i=1}^r {\sinh (z f) \over \sinh (z f_i)}.} This has
a zero at $z=0$ of order $\chi_h + 2g-2$, which is greater than
$\chi_h -3$, for any $g$. Note that, since $f= m_i f_i$, the
factors in the product in \holef\ are polynomials $U_{m_i-1}(\cosh
z f_i)$ where $U_n(x)$ is the Tchebysheff polynomial of the second
kind.

In conclusion, elliptic fibrations are of superconformal simple
type. Notice that the presence of multiple fibers does not affect
the order of vanishing of the SW series. This is a consequence of a
general result that will be proved in section 7.4: the log
transform of a manifold of superconformal simple type is also of
superconformal simple type.

There is a special class of elliptic fibrations that will play an
important role in the examples of section 7.5. These are the simply
connected elliptic fibrations over $\IC P^1$, without multiple
fibers, and with $\chi_h=n$. They are usually denoted by $E(n)$.
One can see that $E(2)$ is a $K3$ surface (it corresponds to
$\chi_h=2$, $g=0$ in the computation above), and $E(1)$ (which has
$b_2^+=1$) is the rational elliptic surface obtained by blowing up
$\IC P^2$ at nine points at the intersection of two different
cubics. The mechanism for the regularity of $F(z)$ is very clear in
these examples: as we increase $\chi_h$ and the degree of
divergence of the prefactor in \expans, new basic classes appear in
such a way that the SW series has a zero with the appropriate order
to compensate the divergence.

In conclusion , we have shown that all minimal complex surfaces are
of superconformal simple type. We consider now the behavior under
blowup, another canonical construction in four-manifold topology.

\subsec{Blowing up preserves superconformal simple type}

The blowup  process is a crucial ingredient in the study of
algebraic geometry and four-manifolds. We study it
here in the smooth category.

The effect of the blowup on the numerical
invariants is to decrease the value of $c_1^2$,  keeping $\chi_h$ fixed, as we
have seen in \vari. By performing an arbitrarily large number of blow-ups
on a four-manifold, we can decrease $c_1^2- \chi_h$ as much as we
want, so if our result is true the SW invariants and the structure
of the basic classes have to change in such a process. We will
prove here the following

\bigskip
\noindent
{\bf Theorem 7.3.1}.  Let $X$ be a manifold of superconformal simple
type. Then, the blownup manifold $\widehat X=X \sharp {\overline
{\IC P^2}}$ is also of superconformal simple type.

\bigskip
{\it Proof}: The proof is rather easy using the known behavior of the SW
invariants under blowup \fsrat. First, recall that manifolds of
superconformal simple type, according to the definition given in
section 6.1, must have $b_2^+>1$ and must be of simple type. The
blowup preserves these two properties, so the statement of the
theorem makes sense (the fact that the blowup of a manifold of
simple type is of simple type is also a result of Fintushel and
Stern \fsrat.) The basic classes of $\widehat X$ are given by $x
\pm E$, where $x$ is a basic class in $X$ (more precisely, the
proper transform of $x$ in $\widehat X$), and $E$ is the
exceptional divisor, with $E^2=-1$. The SW invariants of these
basic classes are given by:
\eqn\swbu{
SW(x\pm E) =SW(x).} By assumption, the holomorphic function
$SW_X^{w_2 (X)} (z)$ \swseries\ has a zero of order $\ge \chi_h (X)
-c_1^2 (X) -3$ at $z=0$. To compute the SW series \swseries\ of the new
manifold $\widehat X$, we have to be careful with $w_2 (\widehat
X)$, because the blowup changes the second Stiefel-Whitney  class.
We can take the integral lifting $\hat c_0 = c_0 + E$, where $c_0$
is an integral lifting of $w_2 (X)$. The computation of the SW
series of the blownup manifold is now straightforward, and we find
\eqn\swsbup{
SW_{\widehat X} ^{w_2 (\widehat X)} (z) = -2 \sinh (z  E) SW_X^{w_2
(X)} (z). } Using \vari, it is immediate to see that the order of
the zero of $SW_{\widehat X} ^{w_2 (\widehat X)} (z)$ is $\ge
\chi_h (\widehat X) -c_1^2 (\widehat X) -3$, {\it i.e.}, $\widehat
X$ is of superconformal simple type. $\spadesuit$

This computation shows that the inclusion of the phase factor in
\swseries\ associated to the second Stiefel-Whitney class is
extremely important. The usual SW series changes under blowup by a
$\cosh E$, and therefore our statement for the order of vanishing
at $z=0$ is simply false for it.

As a corollary of this result, together with the analysis in
sections 7.1 and 7.2, we see that any
complex surface with $b_2^+>1$ is of
superconformal simple type, since any complex surface can be
obtained from a minimal model by a series of blowups.

\subsec{Rational blowdowns}

The blowup process is perhaps the simplest procedure to construct
new manifolds starting from a given one, but we would like to
consider other constructions in order to give more evidence for our
result, and to illustrate some properties of the manifolds of
superconformal simple type. An important construction is the
rational blowdown introduced by Fintushel and Stern \fsrat. This is
a surgical procedure that has three important outcomes: first, it
generalizes in a nice way the usual blowdown. Second, it gives a
very useful description of another classical construction, the log
transform. Finally, one can construct using this procedure
manifolds which lie on the line $c_1^2 =\chi_h -3$, providing in
this way a very nice confirmation of the picture that we have been
developing in this paper. In addition, the rational blowdown will
be one of the ingredients in the construction of the manifolds in
the next section, and our remarks here will also provide a useful
background for these constructions.

\ifig\plumb{This figure, adapted from
Fig. 4.33 of \gompfbook,  illustrates the plumbing construction.
The two disk bundles over $S_1$, $S_2$ become the trivial bundles
$H_1\times D_1^2$, $H_2 \times D_2^2$ after restriction to the disks
$H_1$, $H_2$ in the base manifolds. These two trivial bundles are
then identified after interchanging the factors.}
{\epsfxsize3.0in\epsfbox{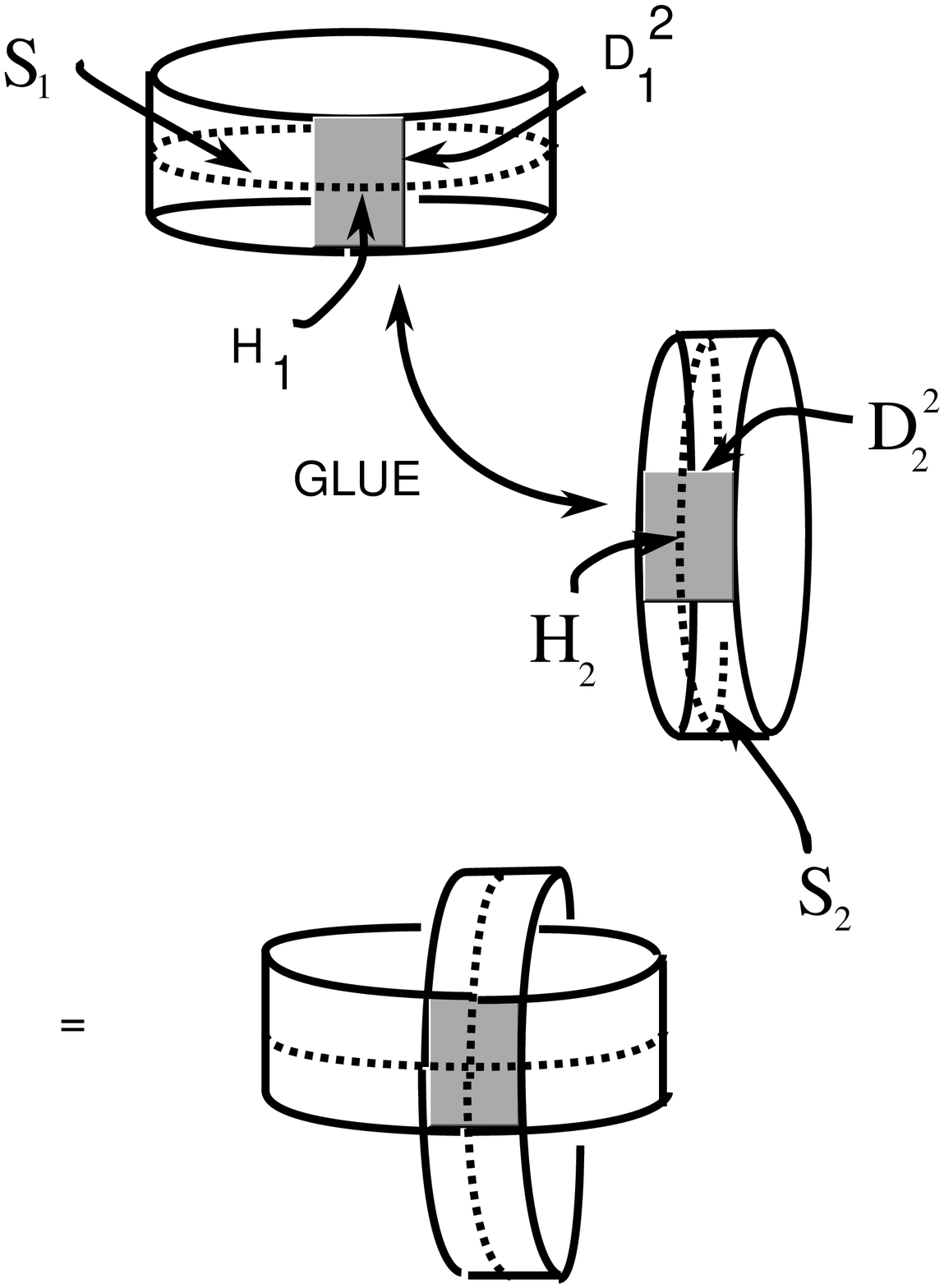}}

To construct a rational blowdown, we must first consider a special
construction in four-manifold topology called ``plumbing" (see
\gompfbook\ for details). Consider two spheres $S_1$, $S_2$
together with disk bundles over them (these are fiber bundles whose
fiber is a disk $D^2$). Restricting the disk bundles to two
nonintersecting
hemispheres $H_1\subset S_1$, $H_2\subset S_2$, (which are also disks), we get
two trivial bundles $H_1\times D_1^2$, $H_2 \times
D_2^2$. The {\it plumbing } of these two configurations consists in
gluing the two total spaces along these trivial bundles, but
interchanging the factors, {\it i.e.} we identify $H_1$ with
$D_2^2$, and $H_2$ with $D_1^2$ (see \plumb\ ).

Consider now $p-1$ disjoint two-spheres,$u_1, u_2, \dots, u_{p-2},
u_{p-1}$ (where $p> 2$) and disk bundles $D_i \rightarrow u_i$
with Euler class $-2$ for $D_1, \dots, D_{p-2}$, and $-(p+2)$
for $D_{p-1}$. If one plumbs these bundles pairwise, following the
sequence $u_1, \dots, u_{p-1}$, one obtains a four-manifold with
boundary which is denoted by $C_p$. According to
\gompfbook, Lemma 8.5.2,  the boundary $\partial C_p$ is the Lens space $L(p^2,
1-p)$. The embedded spheres in $C_p$ satisfy $u_1^2 = \dots =
u_{p-2}^2=-2$, $u_{p-1}^2=-(p+2)$, and they can be oriented in such
a way that $(u_j, u_{j+1})=1$, $1\le j \le p-2$. When $p=2$, the
configuration $C_2$
is just a single two-sphere $u_1$ together with the disk bundle of
Euler number $-4$.

The Lens space $L(p^2, 1-p)$ also bounds another four-manifold
known as the ``rational ball,''
denoted by $B_p$. The rational ball $B_p$ can be constructed as follows \fsrat:
consider the connected sum $\sharp (p-1) \IC P^2$, which has $p-1$
embedded spheres $v_1, v_2, \dots, v_{p-2}, v_{p-1}$ with $p\ge 2$
and such that $v_1^2 = \dots = v_{p-2}^2=2$, $v_{p-1}^2=p+2$. (Again we must
separate cases $p=2$ and $p>2$.)
These spheres can be easily
constructed from the hyperplane divisors in
each copy of $\IC P^2$, and we have
\eqn\hypdiv{
\eqalign{
v_{p-1}&=2h_1 -h_2+ \dots \pm h_{p-1}, \cr
v_{p-2}&=h_1 + h_2,\dots , v_1= h_{p-2}+ h_{p-1},\cr}}
where $h_i$ is the hyperplane divisor of the $i$th copy of
$\IC P^2$. The two-homology classes $v_1, \cdots, v_{p-1}$
are a basis for $H_2 (\sharp (p-1) \IC P^2, \IZ)$. Consider now
a regular neigbourhood of this configuration of $p-1$ spheres. The
complement of
this regular neighborhood in $ \sharp (p-1) \IC P^2$ is
a manifold with boundary $L(p^2, 1-p)$, which is precisely the rational
ball $B_p$.
One can show that the rational ball has $\pi_1 (B_p) = \IZ_{p}$.
Also, $H^2 (B_p, \IQ)=0$, as all the rational two-cohomology
classes of $\sharp (p-1) \IC P^2$ are in the regular neighborhood
of the configuration of $p-1$ spheres.

Let $X$ be a closed four-manifold with an embedded $C_p$ configuration.
The {\it rational blowdown of $X$ along $C_p$},
that will be denoted by $X_p$, is obtained by removing the interior
of $C_p$ and replacing it with $B_p$. Under this operation, the
numerical invariants $\chi_h$, $c_1^2$ change as follows:
\eqn\changerb{
\chi_h (X_p) = \chi_h (X), \,\,\,\,\,\,\,\,\,\ c_1^2 (X_p) =
c_1^2 (X) + p-1.}
We see that the rational blowdown of a configuration $C_2$, which
is associated with a single sphere of self-intersection $(-4)$, has
the same effect on the numerical invariants as a usual blowdown.
This blowdown along $(-4)$-spheres is precisely the construction
that we will use in section  7.7.

Of course, in order to verify our results for rational blowdowns,
we have to relate the basic classes and SW invariants of $X_p$ to
those of the original manifold $X$. There are two general results
that give a partial answer to this problem. Recall that any basic
class is a characteristic element. Fintushel and Stern proved that
the characteristic elements of $X_p$, $\bar x$, come from
characteristic elements of $X$, denoted by $x$. $x$ is called a
lift of $\bar x$. This means that the basic classes of $X_p$ will
be essentially a subset of the set of basic classes of $X$.
Moreover, if $\bar x$ is in fact a basic class of $X_p$, then one
has $SW (\bar x)= SW (x)$. Unfortunately,  this doesn't tell us which $\bar x$
are in fact the basic classes of $X_p$. To clarify this, we need
the details of the embedding $C_p \subset X$. It can
 be shown
that, if $X$ is of simple type, then so is $X_p$ \fsrat. We will
consider in this section two examples of rational blowdowns, where
we are able to obtain a precise description of the resulting basic
classes: the generalized log transform, and the rational blowdown
along configurations $C_{n-2}$ in the elliptic fibrations $E(n)$
(for $n\geq 4$).

\subsubsec{Generalized log transform}

We first need a couple of definitions that will be also useful in
section 7.5. For more details, see for example
\gompfbook\fsknot\fsconst. We say that a smooth four-manifold
contains a {\it cusp neighborhood} if it contains an embedded
submanifold $N$ which is fibered by tori and contains a singular
cusp fiber. An example of manifolds with cusp neighborhoods are the
elliptic fibrations, where the cusp neighborhood can be taken as a
regular neighborhood of a singular cusp fiber together with a
section, and it is often called the nucleus of the elliptic
fibration. A {\it c-embedded torus} is a smoothly embedded torus
$T$, representing a non-trivial homology class $[T]$, which is a
smooth fiber in a cusp neighborhood. In an elliptic fibration, the
generic fiber near a cusp fiber is a c-embedded torus. Notice that
a c-embedded torus has self-intersection zero.

Let $X$ be a smooth four-manifold which contains a cusp
neighborhood, with generic fiber $f$, and consider the blowup $X
\sharp (p-1) {\overline {\IC P^2}}$ along an embedded sphere in the
singular fiber (see \gompfbook, Example 8.5.5. (a), for more details).
It can be shown that the blownup manifold contains
a $C_p$ configuration \fsrat\gompfbook. The rational blowdown of $X
\sharp (p-1) {\overline {\IC P^2}}$ along $C_p$ is, by definition,
the {\it generalized $p$-log transform of $X$}, and will
 be denoted by $X(p)$. The effect of this transform is to create
 a multiple fiber in the fibration, $f_p$, where $f=p f_p$. When
 $X=E(n)$, this construction is equivalent to the usual $p$-log
 transform of Kodaira. Notice that, because of \vari\ and \changerb,
 the generalized $p$-log transform does not change the value of the
 numerical invariants $c_1^2$ and $\chi_h$. The SW invariants and
 basic classes of $X(p)$ can be obtained by combining the results
 on blowup reviewed in the previous section with the results on
 rational blowdowns. The results are as follows. Let $x$ be a
 basic class of $X$. Then, the basic classes of $X(p)$ have the form
\eqn\bcrblow{
x_r = x+ (p-1 -2r) f_p, \,\,\,\,\,\,\,\ r=0, \dots, p-1, } and
$SW(x_r) =SW (x)$.

Using this information, one can compute the SW series \swseries\ of
$X(p)$ in terms of the SW series of $X$. Again, one has to be
careful with the value of $w_2 (X(p))$. A convenient choice of an
integral lifting is $\hat c_0 = c_0 + (p-1)f_p$, where $c_0$ is an
integral lifting of $w_2 (X)$. To compute the SW series, the only
subtlety is the computation of the phase factor involving $w_2
(X(p))$. In order to do that, we need a particular case of the
generalized adjunction inequality proved in \kmad, which gives a
very useful relation between the basic classes and the smooth
topology of four-manifolds. The generalized adjunction inequality
can be stated as follows: let $X$ be a smooth four-manifold with
$b_2^+>1$, and assume that $\Sigma \subset X$ is an embedded,
oriented, connected surface with self-intersection $[\Sigma]^2
\ge0$. For every basic class $x$ of $X$, one has the inequality
\eqn\adj{
2g(\Sigma)-2 \ge [\Sigma]^2+ |(x, \Sigma)|.}
In particular, if $T$ is an embedded torus of self-intersection zero,
one has $([T],x)=0$ for every basic class in $X$. This applies, in
particular, to the generic fiber $f$ in the cusp neighbourhood, hence
$(f, x)=0$ for every basic class. Using this, one obtains:
\eqn\swlog{
SW_{X(p)}^{w_2(X(p))} (z) = (-1)^{((p-1) c_0 \cdot f_p)/2} { \sinh
(z f) \over \sinh (z  f_p)} SW_{X}^{w_2 (X)} (z).} We recognize in
this equation the factor associated to a multiple fiber in the SW
series of an elliptic fibration, given in \holef. Notice that the
order of the zero of $SW_{X}^{w_2 (X)} (z)$ at $z=0$ is not changed
under the $p$-log transform, nor is the value of $c_1^2 (X)- \chi_h
(X)$.  We have then proved the following

\bigskip
\noindent
{\bf Theorem 7.4.1}.  Let $X$ be a manifold of superconformal simple type
that contains a cusp neighborhood. Then, the generalized $p$-log
transform, $X(p)$, is also of superconformal simple type.

\subsubsec{Rational blowdowns of elliptic fibrations}

Once more we summarize some definitions and technical results from \fsrat.
Let $X$ be a manifold of simple type with a $C_p$ configuration. We
say that $C_p$ is {\it tautly embedded} if, for every basic class
$x$ of $X$, one has $(u_i,x)=0$ for $i=1, \dots, p-2$, and
$|(u_{p-1}, x)|\le p$. When the configuration $C_p$ is tautly
embedded, the basic classes of the rational blowdown $X_p$ can be
obtained in a simple way \fsrat. Let $\bar x$ be a basic class of
$X_p$, and let $x$ be a lift in $X$. Then, one must have
\eqn\taut{
|(u_{p-1}, x)|=p.}
Basic classes of $X$ satisfying this are in one-one correspondence
with basic classes $\bar x$ of $X_p$. Moreover, the self-intersection
changes by $\bar x^2 = x^2 + (p-1)$.

To see how this works, consider the elliptic fibrations $E(n)$ with
fiber $f$, and $n \ge 4$. In $E(n)$ one has a sphere $u_{n-3}$
of square $(-n)$,
which is a section of the fibration ({\it i.e.}, $(u_{n-3},f)=1$),
together with $n-4$ spheres $u_i$, $ i=1, \dots, n-4$ with
self-intersection $(-2)$ that are disjoint from the fiber, $(u_i,
f)=0$ for $i=1, \dots, n-4$. One can see that this configuration of
spheres in $E(n)$ is in fact a $C_{n-2}$
configuration \gompfbook\fsrat. The basic classes of $E(n)$ are of
the form $x_r=(n-2-2r)f$, with $r=0, \dots, n-2$. Therefore, the
above configuration $C_{n-2}$ is tautly embedded. Suppose that we
perform a rational blowdown of $E(n)$ along this configuration, to
obtain a manifold that will be denoted by $Y(n)$. On $E(n)$ we have
$|(u_{n-3},x_r)|=|n-2-2r|$, and the only basic classes that satisfy
\taut\ are $\pm (n-2) f$. These two basic classes give the only two
basic classes of $Y(n)$, $\pm \lambda_n$, with $ \lambda_n^2 =n-3$.
Notice that in this process we have ``killed" most of the basic
classes of $E(n)$, to obtain a manifold $Y(n)$ with only one basic
class (in the sense explained in section 6.3). This can be
potentially dangerous for Theorem 4.3.1: It is clear from the
analysis in section 7.2 that, for the elliptic fibrations $E(n)$, we need
all the basic classes $x_r$ to have a zero of the appropriate
order. Remarkably, using \changerb, we find
\eqn\ninvy{
c_1^2 (Y(n))=n-3, \,\,\,\,\,\,\,\,\,\ \chi_h (Y(n))= n,} therefore
$c_1^2 = \chi_h-3$ for this family of manifolds! This is in perfect
agreement with our result in section 6.3, and shows that $Y(n)$ is
also of superconformal simple type. The manifolds $Y(n)$, first
constructed by Fintushel and Stern in \fsrat, therefore
saturate   our
inequality for manifolds with only one basic class \foot{The
elliptic surfaces $E(n)$ contain in general two $C_{n-2}$
configurations, but a further blowdown of the remaining
configuration takes us to the line $c_1^2 = 2\chi_h -6$. In
general, when performing blowdowns along tautly embedded
configurations, the most dangerous possibility for our inequalities
is to perform only one. Further blowdowns increase the value of
$c_1^2$ and maintain the number of basic classes, see the examples
discussed in \fsrat.}.

This ends our remarks about rational blowdowns. It would be
interesting to prove in full generality that the rational blowdown
of a manifold of superconformal simple type is also of
superconformal simple type. We have only been able to prove this
assertion in two important but special cases: generalized log
transforms, and rational blowdowns of $E(n)$ manifolds. This is due
to the fact that, in contrast to the blowup, the change in the
basic classes and SW invariants under a rational blowdown depends
on the particular embedding of the configuration $C_p$. Even for
tautly embedded configurations, a precise knowledge of the
intersection numbers of the embedded spheres $u_i$ with the basic
classes is needed, and is not known in general. In particular, the
intersection form of the manifolds does not behave in a simple
way under the blowdown, because one has in general nonzero intersections
between the embedded spheres and the other two-homology classes in the
manifold.

\subsec{Fiber sum along tori}

The fiber sum is one of the most common procedures to obtain
new four-manifolds starting with simple building blocks, and in fact
all the recent constructions in the geography of four-manifolds are
essentially based on this procedure. Another advantage of the fiber sum is
that, under certain conditions, the SW invariants of the resulting manifold
can be computed
from the invariants
of the original manifolds. In this section, we will review some
of the results concerning this operation and we will prove that
(in some appropriate situations) it preserves the superconformal
simple type condition.

First, we define the fiber sum. Again, see \gompfbook, section 7.1,
for further
details and more precise statements. Let $X_i$, $i=1,2$ be smooth
four-manifolds, and let $F_i \subset X_i$ two-dimensional embedded
surfaces with equal genus $g$ and $[F_i]^2=0$. To construct the
fiber sum of $X_1$ and  $X_2$, we consider tubular neighborhoods of
$F_i$ in $X_i$, $\nu F_i$, and we glue the two manifolds $X_i-\nu
F_i$ along their boundaries using a diffeomorphism $f: F_1
\rightarrow F_2$. The resulting manifold, the fiber sum of $X_1$
and $X_2$ along $F_1$, $F_2$, is denoted by $X_1 \sharp_{F_1=F_2}
X_2$. Notice that, in general, the diffeomorphism type of the sum
depends on the gluing map that we have chosen, but this won't be
the case in the examples we will consider here (essentially because
the fiber sums that we will analyze are along c-embedded tori).
The numerical
invariants of $X_1 \sharp_{F_1=F_2} X_2$ can be computed in terms
of the numerical invariants of $X_i$ and the genus $g$ of $F_i$.
One has \gompfnc:
\eqn\numsum{
\eqalign{
c_1^2 (X_1 \sharp_{F_1=F_2} X_2)&= c_1^2 (X_1) + c_1^2 (X_2) +
8(g-1),\cr
\chi_h (X_1 \sharp_{F_1=F_2} X_2)&= \chi_h (X_1) + \chi_h (X_2) +
(g-1).\cr}}
In the construction that follows, the fiber sum will be performed
along tori, therefore the numerical invariants will simply add
under this operation. Another important aspect of the fiber sum
is that, in appropriate situations, it can be done
on symplectic manifolds in such a way that the symplectic condition is
preserved \gompfnc.

The behavior of the SW invariants under fiber sum along tori has
been analyzed in \mms\fsknot. We will only need   theorems
2.1 and 2.2 in \fsknot. Suppose $X_i$, $i=1,2$
are smooth four-manifolds with $b_2^+>1$, and that they contain
smoothly embedded tori $T_i$ representing nontrivial homology
classes $[T_i]$ with self-intersection $0$. Let $SW_X$ be the
usual SW series
\eqn\useries{
SW_X = \sum_{x} SW (x) {\rm e}^{ x}.} Then, if $X_1 \sharp_{T_1=T_2} X_2$
has $b_2^+>1$, the SW series of the
fiber sum is given by
\foot{This is a kind of gluing formula familiar from axiomatic
approaches to topological field theory. The derivation
from the path integral remains an interesting challenge.}
\eqn\genfiber{
SW_{X_1 \sharp_{T_1=T_2} X_2} = SW_{X_1 \sharp_{T_1=F} E(1)} \cdot
SW_{X_2 \sharp_{T_2=F} E(1)},} where $E(1)$ is the rational
elliptic surface with fiber $[F]$.

 To obtain more concrete results, we
clearly need the expression of each of the factors in the right hand
side of \genfiber. Assume $X$ is as before, and that the torus $T$
is c-embedded. In this case, one has:
\eqn\cemb{
SW_{X \sharp_{T=F} E(1)}= ({\rm e}^{[T]}-{\rm e}^{-[T]}) \cdot
SW_X.} If the torus is not c-embedded, we have to use in principle
other techniques to compute the invariants. We can now prove the
following
\bigskip

\noindent
{\bf Theorem 7.5.1}. Let $X_i$, $i=1,2$ be smooth four-manifolds
with $b_2^+>1$, and
let $T_i$ be c-embedded tori $T_i \subset X_i$.
Let $X=X_1 \sharp_{T_1=T_2} X_2$
be the fiber sum, and suppose that $b_2^+(X)>1$.
If $X_i$ are of superconformal
simple type, then $X$ is also of superconformal
simple type.

\bigskip

\noindent
{\it Proof}: Using \numsum\ for  the fiber sum along tori, one has
\eqn\torinv{
c_1^2 (X)= c_1^2 (X_1) + c_1^2 (X_2),\,\,\,\,\,\,\,\
\chi_h (X)= \chi_h (X_1)
+ \chi_h (X_2).} Notice that, if $X_i$ and $X$ are simply-connected, then
$X$ will have $b_2^+>1$. On the other hand, as a
consequence of a theorem of Morgan,
Mrowka and Szab\'o \mms, any smooth four-manifold
with $b_2^+>1$ and containing a c-embedded torus is of simple type.
Therefore, the statement of our theorem makes sense.
According to \genfiber\ and \cemb,
the SW series \useries\ of $X$ is
given by
\eqn\fibtog{
SW_X = 4 \,(\sinh T)^2 \,SW_{X_1}\cdot  SW_{X_2}.}
Here we denote $[T]= [T_1]=[T_2]$ as a class in $X$.
If we denote by $\{ x^{(1)}_i \}_{i=1, \dots, n_1}$,
$\{ x^{(2)}_j \}_{j=1, \dots, n_2}$ the basic classes of
$X_1$, $X_2$, respectively, \fibtog\ says that the basic classes
of $X$ have the form
\eqn\basfib{
x_{i,j,s} = x^{(1)}_i + x^{(2)}_j + (2-2s) T, }
where $s=0,1,2$, with SW invariant
\eqn\swinv{
SW(x_{i,j,s})=(-1)^s { 2 \choose s} SW(x^{(1)}_i)SW (x^{(2)}_j).}
Notice that $T=T_1=T_2$ has self-intersection $0$, and due to
 the adjunction formula \adj, $(T,x^{(1)}_i)=(T,x^{(2)}_j)=0$
for any $i,j$. As $X$ is of simple type, one has
$(x^{(1)}_i + x^{(2)}_j + (2-2s) T)^2=c_1^2 (X)$, for any $i,j$.
We then see that $(x^{(1)}_i, x^{(2)}_j)=0$, for any $i,j$.
With this information, we can already compute the SW series \swseries.
The simplest choice of integral liftings of $w_2(X_1)$, $w_2 (X_2)$ is
to pick any basic class of $X_1$, $X_2$, say $x^{(1)}_1$, $x^{(2)}_1$.
We also choose the lifting $w_2 (X)=x^{(1)}_1+ x^{(2)}_1$. A simple
computation shows that
\eqn\twistser{
SW_X^{w_2 (X)}(z) = 4 \,(\sinh zT)^2 \,SW_{X_1}^{w_2 (X_1)}
(z) \cdot SW_{X_2}^{w_2 (X_2)} (z).}
With this result, the proof of the theorem is easy. By assumption,
$X_i$ are of superconformal simple type. There are three
different cases to consider:

1) Assume $c_1^2 (X_i)\ge \chi_h (X_i)-3$, for $i=1,2$. Therefore, there are
no constraints on the SW series appearing in the right hand side of
\twistser. According to \torinv, one has $c_1^2 (X)\ge \chi_h (X)-6$, {\it
i.e.}
$\chi_h (X)-c_1^2 (X) -3\le 3$. If we have strict inequality, then
we are done by Proposition 6.1.3, because the order of vanishing of \twistser\
is at least $2$,
due to the factor $(\sinh zT)^2$. If $\chi_h (X)-c_1^2 (X)= 6$,
we must have $c_1^2 (X_i)= \chi_h (X_i)-3$, therefore
$\chi_h(X_i) + \sigma (X_i)$
is odd for $i=1,2$ and the series $SW_{X_i}^{w_2 (X_i)} (z)$ have at least
a zero of order $1$ at $z=0$ (the sum in \sumrules\ with $k=0$ vanishes due the
the symmetry properties of the SW series).
Therefore, the order of the zero at $z=0$ is at least $4>\chi_h (X)-
c_1^2 (X)-3$, and we see that the manifold $X$ is of superconformal simple
type.

2) Assume $c_1^2 (X_1)\ge \chi_h (X_1)-3$, $\chi_h (X_2)-c_1^2 (X_2)-4
\ge  0$. In this case, $SW_{X_2}^{w_2 (X_2)} (z)$ has a zero of order
$\ge \chi_h (X_2)-c_1^2 (X_2) -2$, and $SW_X^{w_2 (X)}(z)$ has a zero of
order $\ge \chi_h (X_2)-c_1^2 (X_2)$. On the other hand, we have by
assumption that $\chi_h (X_2)-c_1^2 (X_2)\ge \chi_h (X)-c_1^2 (X) -3$,
therefore $X$ is of superconformal simple type.

3) Finally, there is the case $\chi_h (X_i)-c_1^2 (X_i)-4
\ge  0$. In this case, $SW_{X_i}^{w_2 (X_i)} (z)$ has a zero
of order $\ge \chi_h (X_i)-c_1^2 (X_i) -2$, therefore
by \twistser,
$SW_X^{w_2 (X)}(z)$ has a zero of order $\ge \chi_h (X)
-c_1^2 (X) -2$. This ends the proof. $\spadesuit$

\subsec{Knot surgery}

The knot surgery construction, introduced by Fintushel and
Stern in \fsknot, is a powerful technique to generate exotic manifolds.
The starting point is a a knot $K$ in ${\bf S}^3$, with (symmetric)
Alexander polynomial
\eqn\alex{
\Delta_K (t)= a_0 + \sum_{j=1}^n a_j (t^j+ t^{-j}).}
If we perform $0$-surgery on $K$, we obtain a three-manifold
$M_K$ with $b_1(M_K)=1$, where the generator of $H_1 (M_K,\IZ)$ is the
meridian $m$ of the knot. In the four-manifold $M_K \times {\bf S}^1$ there
is a smoothly embedded torus $T_m={\bf S}^1 \times m$. Consider now a simply
connected manifold $X$ with $b_2^+>1$ and containing a c-embedded torus $T$.
The {\it knot surgery manifold} $X_K$ is defined as the fiber sum
\eqn\knotsur{
X_K = X \sharp_{T=T_m} (M_K \times {\bf S}^1).}
One can prove that $X_K$ is homeomorphic to $X$, and in particular has
the same values for the numerical invariants. The torus
$T_m$ is not c-embedded in $M_K \times {\bf S}^1$, therefore we can not
use the result \fibtog\ to compute the SW invariants of $X_K$. However, the
SW series \useries\ of $X_K$ is computed in \fsknot\ and is given by
\eqn\knotser{
SW_{X_K} = SW_X \cdot \Delta_K (t),}
where $t={\rm e}^{2T}$. A simple computation, following the
lines in the previous section, shows that
\eqn\knotserw{
SW_{X_K}^{w_2 (X_K)} (z) = SW_X^{w_2 (X)}(z) \cdot
\Delta_K ({\rm e}^{2z T}).}
Therefore, if $X$ is of superconformal simple type, so is $X_K$.

\subsec{An exotic example}

Thus far, we have accumulated some evidence in favour  of our main
result about the regularity of $F(z)$. Moreover, we have seen that
all the complex surfaces are in fact of superconformal simple type,
and we have seen that important constructions in four-manifold
topology (like the blowup, the generalized log transform, the rational
blowdown of $E(n)$, the fiber sum along tori, and the knot surgery)
preserve this condition. In the last few years, such
constructions have been used to find ``exotic" manifolds, like
noncomplex, symplectic manifolds or nonsymplectic, irreducible
manifolds. Since the building blocks of these
constructions are complex surfaces, which are of superconformal
simple type, and the constructions preserve the superconformal
simple type condition, the manifolds constructed in this way
will be of superconformal simple type as well. In this section,
we will present an example of such an exotic family of manifolds
in order to illustrate the interplay between the above constructions
and the superconformal simple type condition.

As we mentioned in section 5.3, the
``wedge" $0< c_1^2 < 2 \chi_h-6$ in the $(\chi_h, c_1^2)$ plane
cannot be filled with minimal complex surfaces, but it can be
filled with minimal (therefore noncomplex) symplectic manifolds.
This wedge is relevant to our results because ``half" of the
manifolds in it are below our line $c_1^2 = \chi_h -3$, and there
are some nontrivial sum rules for the SW invariants that these
manifolds have to satisfy. An interesting example of this construction of a
family of  symplectic manifolds that fills this region is described
in Theorem 10.2.12 of \gompfbook. This construction uses elliptic
fibrations as the building blocks, and the operations of symplectic
fiber sum introduced in \gompfnc, as well as the rational blowdown
discussed in section 7.4.

The starting point of the construction is the elliptic
surface $E(4)$. This surface contains nine disjoint spheres of
square $(-4)$ that are also sections of the fibration. One of these
spheres was used in section 7.4 to perform a rational blowdown
along a configuration $C_2$, but in fact one has nine different
configurations of this type. Consider now $k$ copies of $E(4)_i$, and
perform $b_i \le 8$ rational blowdowns along embedded $(-4)$-spheres
({\it i.e.} $C_2$ configurations) in each $E(4)_i$. The resulting manifolds
will be denoted by $W(b_i)$, $i=1, \cdots, k$. They
were first considered in the Example 5.2 of \gompfnc, and
their Donaldson invariants were computed in \fsrat.

Before proceeding with the construction, we will compute the SW series
\swseries\ of the manifolds $W(b_i)$. To do this, we have to be more
precise about the possible values of $b_i$. Let's denote by
${\alpha}$, ${\beta}$ and ${\gamma}$ the corresponding subsets of
${1, \dots, k}$ with $b_{\alpha}$ odd, $b_{\beta}$ even and
different from zero, and $b_{\gamma}$ zero. We relabel the indices
in such a way that $\alpha=1, \dots, n_1$, $\beta=1, \cdots, n_2$,
and $\gamma=1, \cdots, n_3$, and of course $n_1+n_2+n_3=k$. We also
write,
\eqn\kis{
b_{\alpha} = 2m_{\alpha} +1 , \,\,\,\,\,\ b_{\beta} = 2n_{\beta},
\,\,\,\,\,\ m_{\alpha} \ge 0, \,\ n_{\beta}>0.} Suppose that we
perform $b_i\not=0$ rational blowdowns in the $i$th copy of $E(4)$.
The configurations $C_2$ we are considering are tautly embedded in
$E(4)_i$. According to the results in section
7.4.2, after performing $b_i$ rational blowdowns, we only have
two basic classes $\pm \bar f_i$ in $W(b_i)$, satisfying $\bar f_i^2=b_i$
(each rational blowdown increases the square by one unit, and
we started with $f_i^2=0$). As we have explained,
there are three different cases. If $b_\alpha =2m_\alpha + 1$, and
we choose the lifting $c_0=\bar f_i$, we get
\eqn\alpcase{
SW^{w_2 (W(b_\alpha))}_{W(b_\alpha)}(z)= -2\,\sinh (z \bar f_{\alpha}).}
If $b_\beta = 2n_\beta$, one finds (with the same choice
of lifting)
\eqn\betcase{
SW^{w_2 (W(b_\beta))}_{W(b_\beta)}(z)= 2 \, \cosh (z \bar f_{\beta}).}
Finally, if $b_\gamma=0$,
$W(b_\gamma)=E(4)_\gamma$ and the
SW series is given by the SW series of $E(4)$:
\eqn\firstcase{
SW^{w_2 (W(b_\gamma))}_{W(b_\gamma)}(z)= 4 \,(\sinh (z f_{\gamma}))^2.}
Notice that these manifolds have $c_1^2 (W(b_i))=b_i$,
$\chi_h (W(b_i))=4$, as a consequence of \changerb, and they are of
superconformal
simple type, as $c_1^2 (W(b_i))\ge \chi_h (W(b_i))-3$.

Now, we can construct a family of manifolds filling the wedge
$0< c_1^2 < 2 \chi_h-6$. In $E(4)$ there are two
c-embedded tori disjoint from the nine $C_2$ configurations, $T^{\pm}$
(none of these tori is the fiber of the elliptic fibration, and they are in
fact disjoint from the fiber). Therefore, $T^{\pm}$ are also c-embedded tori
in the rational blowdowns $W(b_i)$. Using these tori,
we can perform the following
fiber sum:
\eqn\efour{
W(b_1) \sharp_{T^+_1=T_2^-} W(b_2) \sharp_{T^+_2=T_3^-} \cdots
W(b_{k-1})\sharp_{T^+_{k-1}=T_k^-} W(b_k),} {\it i.e.} we first sum
$W(b_1)$ and $W(b_2)$ along $T^+_1=T_2^-$. The resulting manifold
still has a $T_2^+$ coming from $W(b_2)$, which we use to sum the
result of the first sum with $W(b_3)$, and so on. We now sum $l$
copies of $(E(1),F)$ to $W(b_k)$ along parallel copies of $T^+_k$,
where $0\le l\le 3$. The resulting manifold will be denoted
by $Y_{b_1, \dots, b_k}$.
Let's first compute the numerical invariants of $Y_{b_1, \dots,
b_k}$. Using the expression \numsum, we obtain:
\eqn\finalnum{
c_1^2 ( Y_{b_1, \dots, b_k}) =b, \,\,\,\,\,\,\,\ \chi_h (Y_{b_1,
\dots, b_k})=4k + l,}
where
\eqn\multiblow{
b=b_1 + b_2 + \dots + b_k.}
Notice that these values realize all the
possible values of $(c_1^2, \chi_h)$ in the wedge: as we can do up
to $8k$  rational blowdowns in total, we can obtain any value of
$b\le 2\chi_h-6=8k+2l-6$. These manifolds are also symplectic,
because both the rational blowdown along $C_2$ configurations
and the fiber sum along tori can be done in such a way that
the symplectic condition is preserved \gompfnc. A simple computation using
\twistser\cemb\alpcase\betcase\ and \firstcase\ shows that
\eqn\seriesy{
\eqalign{
SW_{ Y_ {b_1, \dots, b_k} }^{w_2 (Y_ {b_1, \dots, b_k})}(z) & =
(-1)^{n_1} 2^{3k +l+ n_3 -2} \prod_{\alpha =1}^{n_1} \sinh (z
{\bar f_{\alpha}}) \prod_{\beta =1}^{n_2} \cosh (z {\bar
f_{\beta}})  \prod_{\gamma =1}^{n_3}(\sinh (z {f_{\gamma}}))^2 \cr
& \,\,\,\,\, \cdot (\sinh(z {T_k}))^l \prod_{i=1}^{k-1} (\sinh (z {T_i} ) )^2
.\cr}} Of course, due to the results in section 7.5, the manifold
$Y_ {b_1, \dots, b_k}$ is of superconformal simple type. One can check it
directly by looking at \seriesy. The order of the zero
of this series at $z=0$ is $2k-2+l
+2n_3 + n_1$. On the other hand,
\eqn\lim{
\chi_h -c_1^2 -3 = 4k+l -b-3.}
To see that the manifold $Y_ {b_1, \dots, b_k}$ is of
superconformal simple type, it is enough to prove that the order of
the zero is greater or equal than \lim, or equivalently, that
$b+2n_3 + n_1 -2k +1\ge 0$. Using the definition, $b+2n_3 + n_1 -2k
+1 =b-n_1-2n_2 +1$, but $b\geq n_1 + 2n_2$.
Therefore, $Y_{b_1, \dots, b_k}$ is indeed of superconformal
simple type. This example illustrates very nicely the
delicate balance between the value of $\chi_h -c_1^2 -3$
and the number of zeroes of the SW series.

In the same vein, one can examine many other examples of
noncomplex and even nonsymplectic manifolds in
\gompfnc\szabo\jpark\mszabo, and it is easy to check
that all of them are of
superconformal simple type. In most of the cases, the constructions
use the operations we have analyzed in the previous sections, and
the superconformal simple type condition is automatically satisfied.

\subsec{A conjecture}

As we have seen in section 6, the superconformal simple type
condition is the most natural way to guarantee regularity of $F(z)$
at $z=0$ from the physical point of view. In the last subsections,
we have seen that this condition is also very natural from the
point of view of four-manifold topology: all available four-manifolds
are of superconformal simple type. Based on the evidence we have
reviewed above, we would like to state the following

\bigskip
\noindent
{\bf Conjecture 7.8.1}. Every compact, oriented manifold with $b_2^+>1$
of simple type is of superconformal simple type.
\bigskip

\newsec{Bounds on the number of basic classes}

\subsec{A lower bound and a generalized Noether inequality}

In section 6.3, we found a lower bound for $c_1^2 - \chi_h$ for manifolds
with only one basic class. This bound was a consequence of the sum rules
imposed by our physical theorem 4.3.1. If we assume that the manifold $X$ is
of superconformal simple type,
we can prove a very interesting result: there is
a lower bound on the number of basic classes in terms of $c_1^2 - \chi_h$.
This is a simple consequence of the sum rules \sumrules. In view of our
conjecture 7.8.1, we expect this bound to be true for any manifold with
$b_2^+>1$
and of simple type.

\bigskip
\noindent
{\bf Theorem 8.1.1}. Let $X$ be a four-manifold of superconformal simple type.
If $X$ has $B$ distinct basic classes
(in the sense of section 6.3), and $B>0$, then
\eqn\bound{
B \ge \biggl[ {\chi_h -c_1^2 \over 2} \biggr],}
where $[ \cdot ]$ is the integral part function.
\bigskip

{\it Proof}: To prove this theorem, we will analyze, as usual,
the two different cases $\chi_h + \sigma \equiv 0, \, 1$ mod $2$.

a) If $\chi_h + \sigma$ is even, the sum rules \sumrules\ are
nontrivial when $k$ is even, $k=0,2, \dots, \chi_h -c_1^2 -4$.
Let $x_i$, $i=1, \dots, B$ be the distinct basic classes in
$\CB_X/\{ {\pm1}\}$. Notice that we have modded out by the
involution $x_i \rightarrow -x_i$, therefore no two
basic classes in $\CB_X/\{ {\pm1}\}$ can differ by a sign.
We introduce the notation:
\eqn\factwo{
n_i= 2 SW (x_i) (-1)^{(c_0^2 + c_0 \cdot x_i)/2} ,}
if $x_i \not= 0$, and $n_i=SW(0)(-1)^{c_0^2  /2}$ otherwise. By assumption,
$n_i \not=0$ for  all  $i=1, \dots, B$. To analyze
the sum rules,
we will use the equivalent form   \lastcon, which holds
 for any $S \in H^2 (X, \IC)$. Suppose that the number of
nontrivial equations exceeds the number of basic classes. Then
we have
\eqn\exceed{
{\chi_h - c_1^2 - 4 \over  2} + 1 \geq B.
}
If we consider the first $B$ of these equations we
obtain a linear system of   $B$ equations with $B$ unknowns:
\eqn\fimatrix{
\left( \matrix{ 1& \cdots &1 \cr
(x_1,S)^2 & \cdots & (x_B,S)^2 \cr
\vdots & \ddots& \vdots\cr
(x_1,S)^{2B-2} & \cdots & (x_B,S)^{2B-2}\cr}
\right)
\left(\matrix{ n_1 \cr
\vdots \cr
n_B\cr} \right) = \left(\matrix{ 0 \cr
\vdots \cr
0\cr} \right).}
Since the  $n_i$ are not zero,
 the determinant of the matrix has to be zero.
Therefore,
\eqn\determ{
\prod_{i <j}((x_i,S)^2 - (x_j,S)^2) =\prod_{i<j}(x_i +
x_j,S)\prod_{i<j}(x_i - x_j,S)=0,}
for any $S$.
This determinant is a product of (linear) polynomials in
the coordinates of $S$. A polynomial
ring is a domain, so at least one of the factors in this product must be zero,
{\it i.e.}, there is a pair $i<j$ with   $i,j\in \{1, \cdots, B\}$, so
that
\eqn\alter{
(x_i + x_j,S)=0 \,\ {\rm or} \,\ (x_i - x_j,S)=0}
 for any $S$. Because the
intersection
form is nondegenerate,
this means that $x_i = \pm x_j$ for this pair $i$, $j$. This contradicts the
hypothesis
that the basic classes $x_i$, $x_j$ are in $\CB_X/\{ {\pm1}\}$ (as they
differ by a sign).
Therefore, we must have (taking into account that $\chi_h -
c_1^2$ is even):
\eqn\firstth{
B \ge {\chi_h -c_1^2 \over 2}.}

b) If $\chi_h + \sigma$ is odd, the sum rules apply when $k$ is odd.
Notice that, in this case, $x=0$ cannot be a basic class. Using the
notation
$n_i$ as before, and assuming again that the number of equations is at least
equal to the number of basic classes, $\chi_h -c_1^2 -3 \ge
2B$, we find the system of equations:
\eqn\sematrix{
\left( \matrix{ (x_1,S) & \cdots &(x_B, S) \cr
\vdots & \ddots& \vdots\cr
(x_1,S)^{2B-1} & \cdots & (x_B,S)^{2B-1}\cr}
\right)
\left(\matrix{ n_1 \cr
\vdots \cr
n_B\cr} \right) = \left(\matrix{ 0 \cr
\vdots \cr
0\cr} \right).} The discriminant has to be zero again, hence
\eqn\discri{
\prod_i (x_i,S)\prod_{i<j}(x_i + x_j,S)\prod_{i<j}(x_i - x_j,S)=0.}
As $x_i \not=0$, for any $i=1, \cdots, B$, we find again a contradiction.
This means that
\eqn\secth{
B \ge {\chi_h -c_1^2 -1 \over 2},}
and this ends the proof. $\spadesuit$

\ifig\noether{Lines defining the generalized Noether
inequalities. }{\epsfxsize3.0in\epsfbox{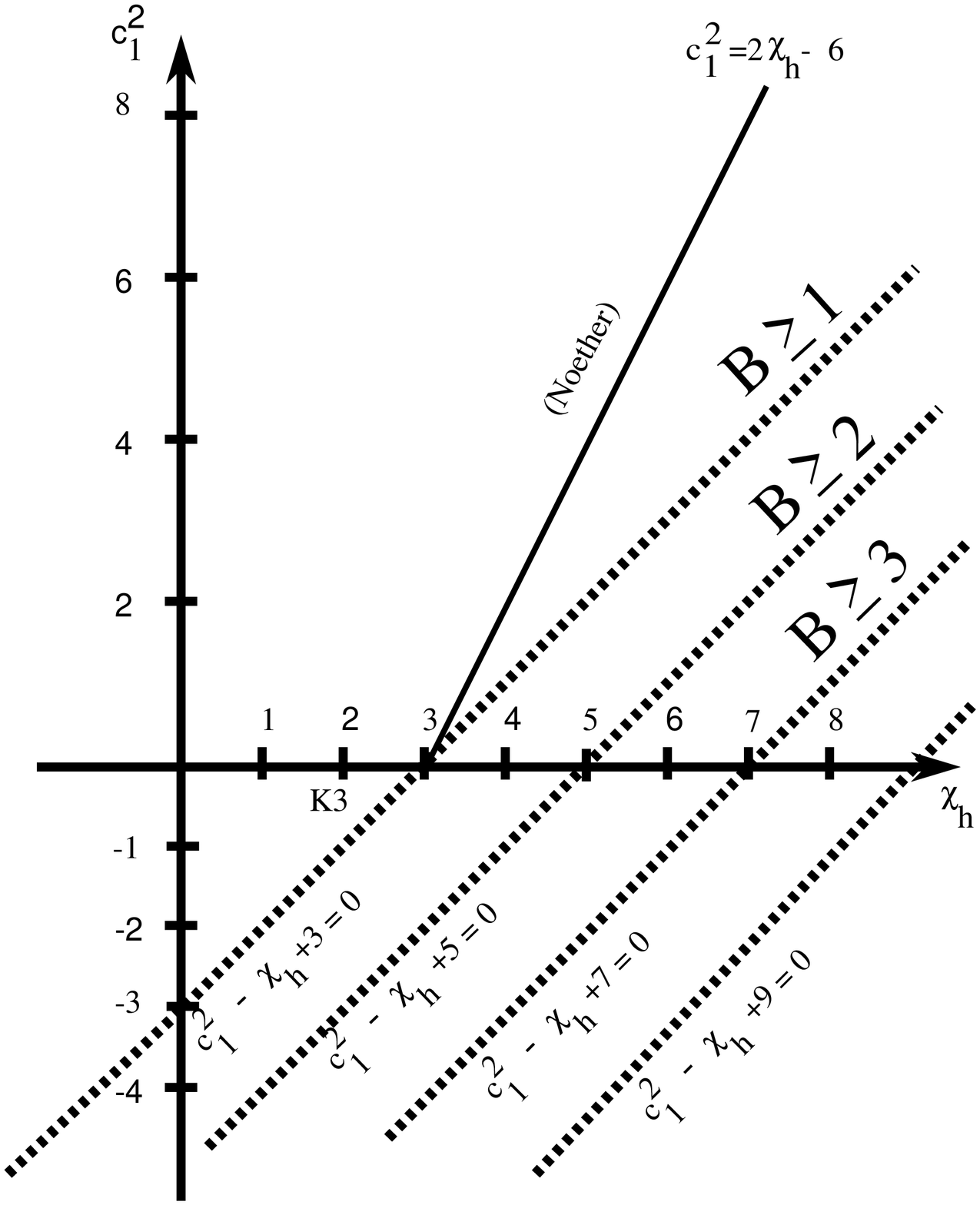}}

The inequality \bound\ is a remarkable fact. It encodes in a simple
way the pattern observed in the previous section, where decreasing the quantity
$c_1^2 - \chi_h$ led to an increase in the number of basic classes. In fact,
theorem 8.1.1 has the following
\bigskip
\noindent
{\bf Corollary 8.1.2.} (Generalized Noether inequality). If $X$ is of
superconformal
simple type and has $B>0$ basic classes, then the following inequality holds
\eqn\noethgen{
c_1^2 \ge \chi_h -2B-1.}
\bigskip

Notice that, if $\chi_h + \sigma$ is even, we cannot have $c_1^2 -\chi_h =
-2B-1$ (as the right hand side is odd), and \noethgen\ is equivalent to
\firstth.
The inequality
\noethgen\ is a generalization of Theorem 6.3.1 under the assumption of the
superconformal simple type condition (if our conjecture is true, this
inequality is valid for any
manifold of simple type with $b_2^+>1$.) It can also be regarded as a
far-reaching generalization of the Noether inequality. We should mention that
the bound \bound\ is in fact
sharp: as it is easy to check, it is saturated by the simply-connected elliptic
fibrations $E(n)$.

\subsec{Upper bounds for the number of basic classes}

It is natural to ask if there are upper bounds for the number of basic classes.
A moment's thought shows that in general
there can be no upper bound depending only
on topological invariants. For example, if we
consider  a manifold with a cusp neighborhood and we perform logarithmic
transforms,
the numerical invariant $c_1^2 -\chi_h$ will remain the same, and at the same
time we will introduce many new basic classes in the manifold (associated to
the multiple fibers.)

On the other hand, if we introduce a metric $g$ on
$X$ then it was already noticed  by Witten in \monopole\ that there is an
upper bound for the number of basic classes depending  on the curvature. We can
combine this with
our lower bound to obtain an interesting corollary in
the theory of Riemannian functionals.

First, let us recall Witten's upper bound. If $2F$ is the
curvature of a Spin$^c$ structure for a solution to the
monopole equations then Witten showed that
\eqn\wittbound{
\eqalign{
{1 \over  4 \pi^2} \int_{X} ( \vert F^+ \vert^2 - \vert F^- \vert^2)
 & ={ c_1^2(X) \over  4}  \cr
\int_{X}  \vert F^+ \vert^2 & \leq {1 \over  16}
\int_X d^4 x \sqrt{g}
(\CR(g))^2 .\cr}
}
Here $\CR(g)$ is the scalar curvature of the metric $g$.
Now, let us choose a basis of harmonic
forms $\omega_\alpha$, $\alpha=1, \dots,
b_2(X) \equiv b$, so that they represent an
integral basis of $H^2(X;\IZ)$. Then we have
\eqn\expdcrv{
\bigl[ 2 { F \over  2 \pi} \bigr] = (2 n_\alpha + c_\alpha)
[ \omega_\alpha ]
}
where $c_0 = [ c_\alpha \omega_\alpha] $, and the $n_\alpha$
are integers. Since harmonic representatives minimize
the $L^2$ norm equation \wittbound\ implies:
\eqn\boundii{
(n_\alpha + \half  c_\alpha) D_{\alpha \beta} (n_\beta + \half  c_\beta) <
\rho^2 \equiv { 1 \over  32 \pi^2} \int_X d^4 x \sqrt{g}
(\CR(g))^2 - { c_1^2(X) \over  4}
}
Here $D_{\alpha \beta} = \int_X \omega_\alpha \wedge * \omega_\beta$ is a
positive definite quadratic form
(depending on $g$) defined on
$H^2(X;\IZ)\otimes \IR$. Equation \boundii\ says that
the basic classes are contained in an ellipsoid of the
form
\eqn\ellipsoid{
\lambda_1^2 y_1^2 + \cdots +\lambda_b^2 y_b^2 \leq \rho^2
} where the $\lambda_i^2$ are the eigenvalues and $y_i$ are
coordinates in the directions of the principal axes of
$D_{\alpha\beta}$. We may order the axes so that $\lambda_1\geq
\lambda_2 \geq \cdots \geq \lambda_b >0 $. (The
$\lambda_i$ depend on the choice of integral basis
$\omega_\alpha$.) Since the $n_\alpha$ are
integers the basic classes can be surrounded by nonoverlapping
$b$-dimensional cubes of volume one in the Euclidean metric $\sum
(dy_\alpha)^2$.  Therefore, the Euclidean volume of any set
completely containing all the cubes surrounding the basic classes
gives an upper bound on the number of classes.  The vertices of any
such cube are vectors of the form $n_\alpha + \half c_\alpha +
e_\alpha$ where $e_\alpha \in \{ \pm \half\}$. Using the Schwarz
inequality and \boundii\ one can show that all such cubes are
completely contained in the ellipsoid of the form \ellipsoid\ with
$\rho \rightarrow \rho + \half \sqrt{b} \lambda_1$. Thus we obtain
the inequality on the number of basic classes:
\eqn\inequality{
2B \leq {\pi^{b/2} \over  \Gamma({b \over  2} + 1) } {( \rho + \half \sqrt{b}
\lambda_1)^b
\over  \sqrt{ \det D_{\alpha\beta}} } .
}
(If $c_1^2=0$ then $x=0$ can be a basic class and
we should replace the LHS of \inequality\ by
$2B-1$.)
When combined with our lower bound, Theorem 8.1,
we find a topological lower bound on the Riemannian functional
in \inequality\ which must hold whenever $X$ supports
basic classes. The maximal eigenvalue
$\lambda_1$ depends on the choice of integral
basis of harmonic forms, while $B$ does not. The
inequality holds for any choice of basis, so we may in
fact take the infimum over the ${\rm SL}(b,\IZ)$ orbit of bases.

\newsec{Conclusions}

In this paper, we have shown that the existence of superconformal points in
gauge theories gives surprising constraints on the topology of four-manifolds.
Superconformal field theory thus
 provides a new tool to study the
relations between the geography problem and gauge invariants. We
have analyzed in detail the simplest realization of this scenario,
the $(1,1)$ point of ${\cal N}=2$, $SU(2)$ supersymmetric gauge
theory with one massive hypermultiplet. Even in this simple
example,
the constraints derived from the analyticity of the
Donaldson-Witten function give  nontrivial information,
revealing   a new line in the $(\chi_h, c_1^2)$
plane, the line $c_1^2 = \chi_h-3$, demarcating two different
kinds of behaviors for the Seiberg-Witten invariants.
In the region below the line the Seiberg-Witten invariants
must satisfy nontrivial sum rules. Using these sum rules
we have proved that any manifold of simple type and
$b_2^+>1$ with only one basic class has to satisfy the
inequality $c_1^2 \ge \chi_h-3$. Our sum rules also
motivate the definition  of manifolds of superconformal simple type, and  in
section seven we have given what we hope
is convincing evidence that all manifolds with $b_2^+>1$ and of
simple type are in fact of superconformal simple type. We have also shown that
the sum rules \sumrules\ encoding the superconformal simple type
condition lead to a   bound for the number of basic classes \bound. This bound
should be regarded
as a generalization of the Noether inequality.

We hope that these results will be also useful in
further exploration of the important open problems such as the
``3/2 conjecture'' and the ``11/8 conjecture.'' In particular, we
have shown that, in the relevant regions, manifolds of $b_2^+>1$ which would
violate these conjectures must satisfy very strong sum
rules on their SW invariants. Indeed, our lower bound
on the number of basic classes, $B$, suggests a new
strategy to approach these problems. If one could establish,
for instance,
an upper bound on $B$ which holds for {\it minimal} manifolds
of $c_1^2<0$ and which violates our bound then
the  ``3/2 conjecture'' would follow. Note that the property of
minimality must play a crucial role in such a hypothetical
upper bound since log transforms on a blowup of
$E(n)$ provide examples of nonminimal manifolds
with $c_1^2<0$ and arbitrarily  large values for $B$.

An interesting consequence of our results is that, as $z
\rightarrow 0$, only a finite number of correlators survive in the
$F(z)$ function, and this number depends on the values of $\chi,
\sigma$. It is tempting to conjecture that the correlation
functions encoded in $F(0)$ are essentially the topological correlation
functions of the superconformal theory. The fact that  only a
finite number of these survive is a strong hint of a selection rule
where the anomalous $U(1)_R$ charge depends on the Euler character
and signature of the four-manifold, as suggested in \gr\ in
a closely related context. This opens the possibility of studying
topological correlators of the still mysterious superconformal
points of ${\cal N}=2$ gauge theories.

The techniques we have used can probably  be considerably
generalized. Superconformal points can be associated with many choices of
vectormultiplet gauge group and hypermultiplet matter representation. In
addition one could consider other topological twists. It should, for example,
be straightforward to generalize our discussion
to certain multicritical points in higher rank theories
with matter. Another  obvious generalization concerns
the higher critical points in $SU(2)$ gauge theory with
$N_f = 2,3,4$. As we have explained, these new critical points involve
noncompact moduli spaces and new kinds of monopole
invariants, so further work needs to be done before interesting
information on 4-manifold topology can be extracted
from these points. But these are matters for future investigation.

\bigskip
\centerline{\bf Acknowledgements}\nobreak
\bigskip

We would like to thank E. Calabi,
D. Freed, T.J. Li and E. Witten for very
helpful discussions and correspondence. We would also like to thank
R. E. Gompf for a very useful discussion on the geography of
four-manifolds. This work  is supported by DOE grant
DE-FG02-92ER40704.

\listrefs

\bye